\documentclass[preprint,12pt,sort&compress]{elsarticle}
\usepackage{amsmath}
\usepackage{mathtools}
\usepackage{amssymb}
\usepackage[mathscr]{eucal}
\usepackage{amscd}
\usepackage{color}
\usepackage{verbatim}
\usepackage{sistyle}
\usepackage{subfigure}
\usepackage{makecell}
\usepackage[usenames,dvipsnames]{xcolor}
 
\usepackage{tikz-cd}
\usepackage{algorithm2e}
\usepackage[permil]{overpic}

\DeclareMathOperator{\SO3}{{{\rm SO(3)}}} 
\DeclareMathOperator{\so3}{{{\rm so(3)}}}

\newcommand{\f}[1]{{\boldsymbol{#1}}}
\newcommand{\baf}[1]{{\bar{\boldsymbol{#1}}}}

\newcommand{\sub}{\subset}
\newcommand{\bEq}{\begin{equation}}
\newcommand{\eEq}{\end{equation}}
\newcommand{\beq}{\begin{equation*}}
\newcommand{\eeq}{\end{equation*}}

\newcommand{\car}{\times}

\newcommand{\emp}{\emph}
\newcommand{\byd}{\,{\raisebox{.092ex}{\rm :}{\rm =}}\,}
\newcommand{\der}{\partial}
\newcommand{\fR}[1]{{\mathbf{#1}}}
\newcommand{\hfR}[1]{\hat{\mathbf{#1}}}

\newcommand{\sepr}[1]{\,\,\,\,\textnormal{{#1}}\,\,\,\,}

\newcommand{\lf}{\left}
\newcommand{\rg}{\right}

\newcommand{\sS}[1]{{\scriptscriptstyle {#1}}}
\newcommand{\Tra}{^{\mathsf{\sS\!T}}}
\newcommand{\Rn}{\text{I\!R}}
\newcommand{\tif}[1]{{\widetilde{\boldsymbol{#1}}}}

\newcommand{\bAl}{\begin{align}}

\newcommand{\ch}[1]{{\check{#1}}}

\newcommand{\ome}{\omega}
\newcommand{\Gam}{\Gamma}
\newcommand{\B}[1]{{\mathbb{#1}}}

\newcommand{\Del}{\Delta}
\newcommand{\vtht}{\vartheta}
\newcommand{\fr}[2]{\frac{#1}{#2}\,}
\newcommand{\seq}{\,\simeq\,}

\newcommand{\wti}[1]{{\widetilde{#1}}}
 
\newcommand{\hf}[1]{\hat{\boldsymbol{#1}}} 
\newcommand{\hti}[1]{\hat{\widetilde{#1}}}

\newcommand{\chf}[1]{{\check{\f{#1}}}}
\newcommand{\alp}{\alpha}

\newcommand{\bdg}{\beq\begin{diagram}}
\newcommand{\edg}{\end{diagram}\eeq}

\newcommand{\Ps}[2]{\prescript{#1}{}{#2}}

\renewcommand{\r}[1]{{\color{black}{#1}}} 

\usepackage{amssymb}
\usepackage{amsthm}

\newproof{prf}{Proof}[section]
\newproof{rmk}{Remark}[section]

\usepackage{lineno}
\usepackage[top=1in,bottom=1in,left=1in,right=1in]{geometry}

\journal{Computer Methods in Applied Mechanics and Engineering}

\begin{document}

\begin{frontmatter}


\title{A fully explicit isogeometric \r{collocation} formulation for the dynamics of geometrically exact beams}

\author[fi]{Giulio Ferri}
\author[tum]{Josef Kiendl}
\author[pv]{Alessandro Reali}
\author[fi]{Enzo Marino\corref{cor1}}
\ead{enzo.marino@unifi.it}
\cortext[cor1]{Corresponding author}

\address[fi]{Department of Civil and Environmental Engineering -- University of Florence, Via di S. Marta 3, 50139 Firenze, Italy.}
\address[tum]{Department of Civil Engineering and Environmental Sciences -- University of the Bundeswehr Munich, Werner-Heisenberg-Weg 39, 85579 Neubiberg, Germany.}
\address[pv]{Department of Civil Engineering and Architecture -- University of Pavia, Via Adolfo Ferrata, 3, 27100 Pavia, Italy.}
\begin{abstract}
We present a fully explicit dynamic formulation for geometrically exact shear-deformable beams. 
The starting point of this work is an existing isogeometric collocation (IGA-C) formulation which is explicit in the strict sense of the time integration algorithm, but still requires a system matrix inversion due to the use of a consistent mass matrix. Moreover, in that work, the efficiency was also limited by an iterative solution scheme needed due to the presence of a nonlinear term in the time-discretized rotational balance equation.
In the present paper, we address these limitations and propose a novel \emp{fully explicit} formulation able to preserve high-order accuracy in space. This is done by extending a predictor--multicorrector approach, originally proposed for standard elastodynamics, to the case of the rotational dynamics of geometrically exact beams. 
The procedure relies on decoupling the Neumann boundary conditions and on a rearrangement and rescaling of the mass matrix. 
We demonstrate that an additional gain in terms of computational cost is obtained by properly removing the \r{angular velocity-dependent} nonlinear term in the rotational balance equation without any significant loss in terms of accuracy.  
The high-order spatial accuracy and the improved efficiency of the proposed formulation compared to the existing one are demonstrated through some numerical experiments covering different combinations of boundary conditions.
\end{abstract}

\begin{keyword}
Isogeometric Analysis \sep Isogeometric Collocation\sep Explicit dynamics \sep Predictor--multicorrector \sep Geometrically exact beams.
\end{keyword}


\end{frontmatter}


\section{Introduction}
In elastodynamics, explicit formulations are often \r{preferred for all those applications where very small time steps are necessary to properly reproduce the complex and fast dynamics of mechanical systems, e.g., under impacts and shock loads \cite{Otto2020,Sun2022}. Thanks to their efficiency and robustness, such methods have been used for example in crash dynamics, metal forming and aerospace simulations. }

Normally, high computational efficiency is pursued through techniques that allow to \r{obtain a diagonal mass matrix} at each time step, \r{such as the row-sum technique and the nodal quadrature method \cite{Wu2005,Elguedj2009,Menouillard2012,Yang2017,Gravenkamp2020}, and to increase the critical time step, e.g., via mass scaling \cite{Voet2023,Yang2017,Gravenkamp2020}.}

\r{The row-sum technique consists in obtaining a diagonal mass matrix from the original one through a summation over the rows. Used in conjunction with shape functions that form a partition-of-unity, it preserves the total mass \cite{Gravenkamp2020}. 
It is easy to implement, allows to increase the critical time step, but permits achieving only second-order accurate frequencies~\cite{Cottrell_etal2006}.
Moreover, row-sum technique may fail, leading to singular or indefinite lumped mass matrices. Such a drawback depends on the spatial discretization scheme employed \cite{Gravenkamp2020}, and it is prevented by non-negative partition of unity methods \cite{Voet2023}.}  
The nodal quadrature method relies on a special choice of quadrature nodes. As noted in \cite{Voet2023}, it may preserve high-order accuracy at the cost of deteriorating its efficiency. 
Lastly, mass scaling schemes still require the matrix inversion, but they improve the efficiency of explicit methods by increasing the critical time step \cite{Tkachuk2015,Schaeuble2017}. 
A comparison of these approaches in the context of Spectral Element Method is presented in \cite{Duczek2019}. 

In isogeometric analysis (IGA)~\cite{Hughes2005,Cottrell_etal2006}, the local support of NURBS and B-Splines \cite{Piegl1997,Cottrell2009} makes the consistent mass matrix banded with a bandwidth determined by the basis function degree. Standard lumping approaches turn out to deliver at most second-order accurate results \cite{Anitescu2019,Li2022}. Therefore, finding lumping procedures capable of preserving high-order spatial accuracy is fundamental to exploit the full potentialities of IGA formulations in explicit dynamics. A basis function transformation is proposed in \cite{Li2022} and adopted for IGA structural vibration analysis, proving an improved frequency accuracy. A ``dual lumping'' procedure in the context of Petrov-Galerkin IGA methods is proposed in \cite{Anitescu2019}. It is shown via numerical examples on 2D domains that the method is superior to standard row-sum techniques and achieves a similar accuracy with respect to consistent matrix formulations. A similar approach is adopted in \cite{Nguyen2023} for the spectral analysis of beams, plates and shells retrieving the higher-order accuracy of consistent mass matrix formulations. 

\r{Within} the IGA framework, to achieve higher efficiency levels keeping the attributes of classical IGA, the isogeometric collocation (IGA-C) method is proposed in~\cite{Auricchio2010,Auricchio2012}. IGA-C is based on the discretization of the strong form of the governing equations and requires one evaluation (collocation) point per degree of freedom. Compared to Galerkin-based IGA and Galerkin-based FEA, IGA-C can be orders of magnitude faster~\cite{Schillinger2013} to achieve a specified level of accuracy. Moreover, IGA-C naturally circumvents the known problem of sub-optimal quadrature rules in weak-form IGA~\cite{Adam_etal2015,Fahrendorf_etal2018,Sangalli&Tani2018}.

IGA-C-based methods have been successfully applied to a wide range of problems, including elasticity, hyperelasticity, and elastoplasticity~\cite{Auricchio2010,Auricchio2012,Schillinger2013,Kruse2015,Fahrendorf_etal2020}; 
phase-field~\cite{Gomez2014,Schillinger_etal2015,Fedeli_etal2019}; contact \cite{DeLorenzis2015,Kruse2015,Weeger_etal2017a}; 
linear beams~\cite{BeiraodaVeiga2012,Auricchio2013,Kiendl2015a,KiendlAuricchioReali2017,Balduzzi_etal2017,Reali2015,Marino_etal2020,Ignesti2023};
nonlinear planar beams~\cite{Maurin_etal2018b}; 
plates and shells~\cite{Reali2015,Kiendl2015,KiendlMarinoDeLorenzis2017,Maurin_elal2018}; 
electromechanics problems~\cite{Torre_etal2023}; 
\r{geometrically} exact static   
\cite{Marino2016a,Weeger_etal2017a,Marino2017b,Ferri2023} and dynamic \cite{Weeger2017,Marino2019a,Marino2019b} beams.

Due to its efficiency, IGA-C is particularly attractive for explicit dynamics. For a two-dimensional linear elastodynamic problem, it is conveniently used in combination with a predictor--multicorrector algorithm that allows to \r{obtain a diagonal mass matrix} \cite{Auricchio2012}.
For the same problem, this approach is further developed in~\cite{Evans_etal2018}, where an explicit higher-order space- and time-accurate scheme is proposed. Higher-order time accuracy is achieved through explicit Runge-Kutta methods. 

Among the existing IGA-C formulations for the problem of geometrically exact beams~\cite{Marino2016a,Marino2017b,Weeger2017,Weeger_etal2017a,Marino2019a,Marino2019b,Ferri2023,Marino_and_Ferri2024}, only in \cite{Marino2019a} an explicit scheme is proposed. In that work, the exceptionally well-performing $\SO3$-consistent explicit time integrator for rigid body dynamics \cite{Krysl_and_Endres2005} is extended to the rotational dynamics of beams. However, following the distinction made in \cite{Evans_etal2018}, the method in~\cite{Marino2019a} is considered explicit more in the applied mathematics sense. Since it employs a consistent mass matrix, the formulation still requires the mass matrix inversion at each time step. Moreover, it needs a Newton-Raphson scheme for the solution of the entire system of equation due to a nonlinear term appearing in the time-discretized rotational balance equation. 

In the present paper, we address both these issues  proposing a novel formulation able to preserve the high-order accuracy in space without the need for any matrix inversion. We refer to this formulation as \emp{fully explicit}. 
We extend the predictor--multicorrector approach of \cite{Evans_etal2018} to geometrically exact beams demonstrating the capability to achieve an unprecedented level of efficiency keeping al attributes in terms of accuracy in space. The proposed lumping procedure relies on two main actions: i) decoupling the translational and rotational equations in the Neumann boundary conditions, which are enforced without any treatment, such as hybrid collocation-Galerkin or enhanced collocation~\cite{DeLorenzis2015}; and ii) rearranging and rescaling of the mass matrix in a convenient form. 
Moreover, an extra efficiency gain is obtained by removing the \r{angular velocity-dependent} nonlinear term appearing in the rotational balance equation, bypassing the need for a time-consuming iterative scheme.

Robustness and efficiency of the proposed formulation are proved through demanding numerical tests, covering different combinations of boundary conditions.

The paper is structured as follows: in Section 2, the IGA-C explicit scheme for the nonlinear dynamics of shear-deformable beams is briefly recalled. In Section 3, we present the fully explicit IGA-C formulation, focusing on the extension of the predictor--multicorrector approach to the dynamics of geometrically exact beams. In Section 4, we assess the performance of the proposed formulation through some numerical experiments. Finally, the main conclusions of our work are drawn in Section 5.

\section{A brief review of the IGA-C explicit scheme for beam dynamics}
In this section, we briefly review the explicit scheme proposed in \cite{Marino2019a} for the dynamics of spatial shear-deformable beams undergoing finite motions. Firstly, time and space discretizations of the governing equations and the $\SO3$-consistent configuration update are introduced. Then, some critical aspects related to the solution procedure are discussed. 

\subsection{Time discretized governing equations in local form}
The strong form of the translational and rotational balance equations for geometrically exact shear-deformable beams \cite{Simo1985} can be rewritten in terms of kinematic quantities exploiting the linear elastic constitutive equations

\bEq\label{eq:NM_material}
\f n = \fR R\B C_N \f\Gam_N  \sepr{and} \f m = \fR R\B C_M \f K_M\,,
\eEq
where $\f n$ and $\f m$ are the internal forces and moments; $\B C_N =\text{diag}(GA_1,EA,GA_3)$ and $\B C_M = \text{diag}(EJ_1,GJ,EJ_3)$ are the elasticity tensors. $\f\Gam_N=\fR R\Tra \f c,_s - \fR R_0\Tra \fR c_0,_s$ and 
$\f K_M =\rm axial (\tif K - \tif K_0) = \f K - \f K_0$, are the material strain measures of the beam. $\f c$ and $\f c_0$, both belonging to $\Rn^3$, denote the beam centroid in the current and initial configuration, respectively.
$\fR R$ and $\fR R_0$, both belonging to $\SO3$, are the orthogonal operators that identify the spatial (rigid) rotation of the beam cross sections. We omit that, in general, the involved quantities (excepting the elasticity tensors) are parameterized over time $t\in[0,T]\sub \Rn$ and space $s\in [0,L]\sub \Rn$, where $T$ is the length of the time domain and $L$ is the length of the beam centroid line in the reference configuration. $\tif K\sepr{and} \tif K_0\in\so3$ are the current and initial curvature tensors in the material form. Quantities with subscript ``$0$'' refer to the initial configuration, therefore they are only space-depended. With $(\cdot),_s$ we express the derivative with respect to the abscissa $s$.

Substituting Eq.~\eqref{eq:NM_material} into the well known local form of the governing equations, see for example \cite[Eqs. 3.3a and  3.3b]{Simo1985}, and discretizing it in time lead to  
\bAl
\mu \f a^n & = \fR R^n \tif K^n \B C_N \f\Gam_N^n + \fR R^n \B C_N\f\Gam_{N,s}^n +  \baf n^n  \label{eq:ge_fkin_dyn_tdisc}\,,\\ 
\f j^n \f \alp^n + \tif \ome^n \f j^n \f\ome^n  & = \fR R^n \tif K^n \B C_M \f K_M^n  + \fR R^n \B C_M\f K_{M,s}^n + \f c,_{s}^n \car \fR R^n\B C_N\f\Gam_N^n + \baf m^n\,, \label{eq:ge_mkin_dyn_tdisc}
\end{align}
where $(\cdot)^n$ denotes any quantity evaluated at time $t = t^n$. 
Indicating with $\dot{(\cdot)}$ the derivative with respect to time, $\f a=\dot{\f v}$ and $\f v=\dot{\f c}$ are the spatial acceleration and velocity of the beam centroid, while $\f\alp=\dot{\f\ome}$ and $\f\ome$ are the spatial angular acceleration and velocity vectors of the beam cross section. 
$\f \ome$ is the axial vector\footnote{With $\wti{(\cdot)}$ we indicate elements of $\so3$, that is the set of $3 \times 3$ skew-symmetric matrices. In this context, they are used to represent angular accelerations, curvature matrices, and infinitesimal incremental rotations. 
For any skew-symmetric matrix $\tif a\in\so3$,  $\f a = \textnormal{axial}(\tif a)$ indicates the axial vector of $\tif a$ such that $\tif a \f h= \f a \car \f h$, for any $\f h\in \Rn^3$, where $\car$ is the cross product.} of the skew-symmetric tensor $\tif \ome \byd \dot{\fR R}\fR {R}\Tra$; 
$\f j=\fR R \f J \fR R\Tra$; is the spatial inertia tensor, while $\mu$ is the mass per unit length; 
$\bar{\f n}$ and $\bar{\f m}$ are the distributed external forces and moments per unit length. 

The set of governing equations is completed by the boundary and initial conditions, given in the spatial form as
\bAl
\f \eta   & =\baf \eta_c \sepr{or} \f n = \baf n_c   \sepr{with}	s = \{0,L\}\,, t\in[0,T]\,, \label{eq:bcseta}\\
\f \vtht  & =\baf \vtht_c \sepr{or} \f m  = \baf m_c \sepr{with}	s = \{0,L\}\,, t\in[0,T]\,,   \label{eq:bcstht}\\
\f v   &=\f v_0    \sepr{with}  s\in(0,L) \sepr{and} t = 0\,,\label{eq:icv}\\
\f \ome   &=\f \ome_0   \sepr{with}  s\in(0,L) \sepr{and} t = 0\,.\label{eq:icome}
\end{align}
where $\baf \eta_c$ and $\baf \vtht_c$ are the prescribed displacements and rotations of the beam ends, $\baf n_c$ and $\baf m_c$ are the external concentrated forces and couples, while $\f v_0 $ and $\f \ome_0$ are the initial velocities and angular velocities of the beam, respectively.
  
\subsection{Consistent update of the right hand sides of the governing equations}
The beam configuration is fully determined by the pair $(\f c, \fR R)$ for any $t\in[0,T]$ and $s\in[0,L]$. 
In a time-discretized context, the geometrically consistent update of the beam configuration \cite{Marino2016a,Marino2019b}, 
say $(\f c^{(n-1)},\fR R^{(n-1)}) \to (\f c^n, \fR R^{n})$, is made as follows 
\bAl
\f c^{n} & =   \f c^{(n-1)} + \f\eta^{(n-1)}\,,\label{eq:update_c}\\
\fR R^{n} &= \exp(\tif \vtht^{(n-1)}) \fR R^{(n-1)}\label{eq:update_R}\,,
\end{align}
where $\f\eta^{(n-1)}\in \Rn^3$ and $\tif \vtht^{(n-1)}\in \so3$ denote the incremental displacement and spatial rotation at $s$, respectively. $\f\eta^{(n-1)}$ acts through a standard translation (additive rule) of the beam centroid $\f c^{(n-1)}$, whereas $\tif \vtht^{(n-1)}$ acts through the (multiplicative) group composition rule, being  $\exp(\tif \vtht^{(n-1)})\in\SO3$ the incremental rotation superimposed to the current rotation $\fR R^{(n-1)}$. $\exp : \so3 \to \SO3$ is the exponential map of the rotation group which is known in closed form (Rodrigues formula) \cite{Argyris1982}. 
The above geometrically consistent updating formulas rely on the proper construction of the tangent space to the configuration manifold whose details can be found in \cite{Simo&Vu-Quoc1988,Makinen2007,Makinen2008,Marino2019b}.

\subsection{IGA dicretization and existing solution method}
Following the IGA paradigm, the beam centroid $\f c$, along with the displacements and rotations $\f\eta$ and $\f\vtht$, 
velocities $\f v$ and $\f\ome$, and accelerations $\f a$ and $\f\alp$ are discretized in space as follows 
\bAl
\f c (u) & \seq \sum_{j= 0}^{\rm n} R_{j,p}(u) \chf c_j\,,\label{eq:cu}\\
\f\vtht (u) & \seq \sum_{j= 0}^{\rm n} R_{j,p}(u) \chf\vtht_j\,,\label{eq:thtu}\\
\f\eta (u) & \seq \sum_{j= 0}^{\rm n} R_{j,p}(u)\chf\eta_j\,,\label{eq:etau}
\end{align}
\bAl
\f\ome (u) & \seq \sum_{j= 0}^{\rm n} R_{j,p}(u) \chf\ome_j\,,\label{eq:omeu}\\
\f v (u) & \seq \sum_{j= 0}^{\rm n} R_{j,p}(u)\chf v_j\,,\label{eq:vu}
\end{align}
\bAl
\f\alp (u) & \seq \sum_{j= 0}^{\rm n} R_{j,p}(u) \chf\alp_j\,,\label{eq:alpu}\\
\f a (u) & \seq \sum_{j= 0}^{\rm n} R_{j,p}(u)\chf a_j\,,\label{eq:au}
\end{align}
where $\ch{(\cdot)}_j$ is the $j$th control value of the related field and 
$R_{j,p}$ is the $j$th NURBS basis function of degree $p$ depending on the parametric abscissa $u\in [0,\,1]$ \cite{Piegl1997,Cottrell2009}.

In the present formulation, among the above control quantities, the only unknowns are $\chf\alp_j$ and $\chf a_j$. The remaining control quantities, $\chf\eta_j$, $\chf\vtht_j$,$\chf v_j$, and $\chf\ome_j$, are computed with the following $\SO3$-consistent explicit \r{central difference scheme} \cite{Krysl_and_Endres2005}
\bAl
\chf \eta_j^{(n-1)} & = h \chf v_j^{(n-1)} + \fr {h^2}2\chf a_j^{(n-1)}\,,\sepr{with} j = 0,\ldots,\rm n\,, \label{eq:eta_nm1}\\
\chf\vtht_j^{(n-1)} & = h \chf\ome_j^{(n-1)} + \fr {h^2}2\chf\alp_j^{(n-1)}\,,\sepr{with} j = 0,\ldots,\rm n\,. \label{eq:tht_nm1}
\end{align}
\bAl
\chf v_j^n & = \chf v_j^{(n-1)} + \fr h 2 \lf( \chf a_j^{(n-1)} +\chf a_j^n\rg) = \chf v_{pj}^{(n-1)} + \fr h 2\chf a_j^n  \,, \label{eq:vn_update}\\
\chf \ome_j^n & =\chf \ome_j^{(n-1)} + \fr h 2 \lf(\chf \alp_j^{(n-1)} +\chf\alp_j^n\rg) = \chf \ome_{pj}^{(n-1)} + \fr h 2 \chf\alp_j^n\,, \label{eq:omen_update}
\end{align}
where we have defined $\chf v_{pj}^{(n-1)} = \chf v_j^{(n-1)} + \fr h 2 \chf a_j^{(n-1)}$ and $\chf \ome_{pj}^{(n-1)} = \chf \ome_j^{(n-1)} + \fr h 2 \chf \alp_j^{(n-1)}\,$. $h$ is the time step size. 
The above scheme allows to express the right hand side of both Eqs.~\eqref{eq:ge_fkin_dyn_tdisc} and~\eqref{eq:ge_mkin_dyn_tdisc} in terms of  quantities known from previous time steps.

The balance equations are collocated at the standard Greville abscissae $u^c_i$ with $i = 1,\ldots, \rm n$ \cite{Auricchio2010} as follows\footnote{To simplify the notation, collocated quantities at $u = u^c_i$ are denoted by $(\cdot)_i$.} 
\bAl
\mu \f a_i^n & = \f \psi^n_i\, \sepr{with} i = 1,\ldots,\rm n-1  \label{eq:ge_fkin_dyn_tdisc2}\,,\\ 
\f j_i^n \f \alp_i^n + \tif \ome_i^n \f j_i^n \f\ome_i^n  & = \f \chi^n_i \, \sepr{with} i = 1,\ldots,\rm n-1\,, \label{eq:ge_mkin_dyn_tdisc2}
\end{align}
where the collocated right-hand side terms, known form the previous time step, have been defined as follows  
\bAl
\f \psi^n_i  & = \lf[\fR R^n \tif K^n \B C_N \f\Gam_N^n + \fR R^n \B C_N\f\Gam_{N,s}^n +  \baf n^n\rg]_i\,,\label{eq:psini}\\
\f \chi^n_i & = \lf[\fR R^n \tif K^n \B C_M \f K_M^n  + \fR R^n \B C_M\f K_{M,s}^n + \f c,_{s}^n \car \fR R^n\B C_N\f\Gam_N^n + \baf m^n\rg]_i\,.\label{eq:chini}
\end{align}

It is noted that substituting Eq.~\eqref{eq:omen_update} into Eq.~\eqref{eq:ge_mkin_dyn_tdisc2} leads to the following nonlinear rotational balance equation
 \bEq
 \f j_i^n \f \alp_i^n + [\f \ome^{(n-1)}_{p,i}+\fr{h}{2}\f \alp_i^n]\times \f j_i^n [\f \ome^{(n-1)}_{p,i}+\fr{h}{2}\f \alp_i^n]  = \f \chi^n_i \, \sepr{with} i = 1,\ldots,\rm n-1\,. \label{eq:ge_mkin_dyn_tdisc3}
\eEq
 
The above nonlinear term in $\f\alp^n_i$ necessitates a Newton-Raphson scheme with a tangent operator given by \cite{Krysl_and_Endres2005,Marino2019b}
\bEq
\fr{\der\fR r_i^n(\hf\alp^n_i)}{\der \f\alp^n_i}\sum_{j= 0}^{\rm n} R_{j,p}{\Del\chf \alp^n_j} = {-\hfR r_i^n} \, \sepr{with} i = 1,\ldots,\rm n-1\,,
\eEq
where 
\bEq
{\fR r_i^n}  = \f j_i^n \f \alp_i^n + [\f \ome^{(n-1)}_{p,i}+\fr{h}{2}\f \alp_i^n]\times \f j_i^n [\f \ome^{(n-1)}_{p,i}+\fr{h}{2}\f \alp_i^n]  - \f \chi^n_i  \, \sepr{with} i = 1,\ldots,\rm n-1\,.
\eEq

The boundary equations can also be expressed in terms of primary unknowns. For the Dirichlet boundary conditions, assuming a clamped end and exploiting Eqs.~\eqref{eq:vn_update} and \eqref{eq:omen_update}, we have
\bAl
\chf a^n_j  & = - \fr{1}{h} \chf v^{(n-1)}_{pj}\,,\\
\chf \alp^n_j  & = - \fr{1}{h} \chf \ome^{(n-1)}_{pj}\,. 
\end{align}

Similarly, the Neumann boundary conditions become
\begin{gather}
\!\!\Ps{1}{\f\psi}^n_i h^2   \sum_{j = 0}^{\rm n} R_{j,p} \chf\alp^n_j +  \Ps{2}{\f\psi}^n_i h^2 \sum_{j = 0}^{\rm n} R'_{j,p} \chf a^n_j\!=\! \baf\psi^n_i - h \lf(\Ps{1}{\f\psi}^n_i \sum_{j = 0}^{\rm n} R_{j,p} \chf\ome^{(n-1)}_{pj} + \Ps{2}{\f\psi}^n_i \sum_{j = 0}^{\rm n} R'_{j,p}\chf v^{(n-1)}_{pj}\rg),	\label{eq:coll_disc_tran_Neubc2}\\
h^2 \lf(\Ps{1}{\f\chi}^n_i  \sum_{j = 0}^{\rm n} R_{j,p} + \Ps{2}{\f\chi}^n_i  \sum_{j = 0}^{n} R'_{j,p}\rg) \chf\alp^n_j\!=\!
\baf\chi^n_i - h \lf(\Ps{1}{\f\chi}^n_i  \sum_{j = 0}^{\rm n} R_{j,p} + \Ps{2}{\f\chi}^n_i  \sum_{j = 0}^{\rm n} R'_{j,p}\rg) \chf\ome^{(n-1)}_{pj}  \,,\label{eq:coll_disc_rot_Neubc2}
\end{gather}
where we have set 
\bAl
\Ps{1}{\f\psi}^n_i & = \lf[\hfR R^n \B C_N \hfR R\Tra{^n} \hti{\f c},_s^n - \wti{\lf(\hfR R^n \B C_N \hf\Gam^n_N\rg)} \rg]_i\,, \label{eq:1psini}\\
\Ps{2}{\f\psi}^n_i & =  \lf[ \hfR R^n\B C_N \hfR R\Tra{^n} \rg]_i\,, \\
\Ps{1}{\f\chi}^n_i & =\lf[- \wti{\lf(\hfR R^n \B C_M \hf K^n_M\rg)} \rg]_i\,, \\
\Ps{2}{\f\chi}^n_i & =\lf[ \hfR R^n\B C_M \hfR R\Tra{^n} \rg]_i\,, \\
\baf\psi^n_i  & =  - \lf( \hfR R^{n} \B C_N\hf\Gam_N^{n} - \baf n_c^{n}\rg)_i\,,\\
\baf\chi^n_i  & =  - \lf( \hfR R^{n} \B C_M\hf K_M^{n} - \baf m_c^{n}\rg)_i\,,\label{eq:bafchini}
\end{align}
with the collocation point that can be either $i=0$ or $i=\rm n$, depending on which end of the beam the condition holds.

As already noted above, a high efficiency of the existing explicit IGA-C solution method \cite{Marino2019a} is prevented by two main reasons: i) the use of a consistent mass matrix; 
ii) the need for a Newton-Raphson scheme for the solution of the entire system of equation, which is made nonlinear by the time discretized rotational balance equation.

In the following Section we address these issues proposing a \emp{fully explicit} solution method.

\section{Fully Explicit IGA-C solution method}
The consistent mass matrix problem is addressed by extending the predictor--multicorrector approach proposed in \cite{Auricchio2012,Evans_etal2018} to the nonlinear rotational dynamics. To do that, first a decoupling of the Neumann boundary equations is necessary. Second, we bypass the Newton-Raphson algorithm assuming an upfront linearized form of the rotational balance equation.

\subsection{The predictor--multicorrector approach for rotational dynamics}\label{PMA}
At each time step, a system in the general form $\f M \f x=\f b$, where $\f M$ is the mass matrix, $\f x$ is the vector of unknowns, and $\f b$ is the force vector, must be solved. 
If $\f M$ is diagonal, the system is straightforwardly solved without any matrix inversion. If not, a lumping procedure should be adopted to promote efficiency. 

In its original form \cite{Auricchio2012,Evans_etal2018}, the predictor--multicorrector method allows to exploit a lumping of the mass matrix through the following iterative scheme
\begin{equation}
\left\{
\begin{alignedat}{3}
&\f x^{0}  = \f 0 \\
& \textnormal{for} ~i=0,...,r-1\\
&\hspace{18pt}\f M_L \Delta\f x^i=\f b-\f M \f x^i\\
&\hspace{18pt}\f x^{i+1}  = \f x^{i} +\Delta\f x^i \\
&\rm{end}
\end{alignedat}
\right.
\end{equation}
where $\f M_L $ is the lumped mass matrix, which coincides with the identity matrix, $\f I$, and $r$ denotes the number of corrector passes. Convergence is guaranteed if  $\rho(\f{M-I})<1$, where  $\rho(\f{M-I})$ is the spectral radius of the iteration matrix. 

To apply the above algorithm in a (finite) rotational beam dynamic context, we need to recast the banded mass and inertia matrices such that the spectral radius condition is fulfilled. To do that, we first need to decouple the Neumann boundary equations by making the assumption that $\sum_{j = 0}^{\rm n} R_{j,p}\chf\alp^{n}_j=\sum_{j = 0}^{\rm n} R_{j,p}\chf\alp^{(n-1)}_j$. This allows to move from the left-hand side to the right-hand side the first term in Eq.~\eqref{eq:1psini}. Note that this term is multiplied by $h^2$, therefore, considering that in explicit dynamics the time steps are normally very small, we expect no significant loss of accuracy.
The Neumann boundary conditions then become
\bEq
\Ps{2}{\f\psi}^n_i h^2 \sum_{j = 0}^{\rm n} R'_{j,p} \chf a^n_j= \baf\psi^n_i - h (\Ps{1}{\f\psi}^n_i \sum_{j = 0}^{\rm n} R_{j,p} \chf\ome^{(n-1)}_{pj} + \Ps{2}{\f\psi}^n_i \sum_{j = 0}^{\rm n} R'_{j,p}\chf v^{(n-1)}_{pj})-\Ps{1}{\f\psi}^n_i h^2   \sum_{j = 0}^{\rm n} R_{j,p} \chf\alp^{(n-1)}_j,	\label{eq:coll_disc_translat}\\
\eEq
\bEq
\Ps{2}{\f\chi}^n_i  h^2\sum_{j = 0}^{n} R'_{j,p} \chf\alp^n_j=
\baf\chi^n_i - h (\Ps{1}{\f\chi}^n_i  \sum_{j = 0}^{\rm n} R_{j,p} + \Ps{2}{\f\chi}^n_i  \sum_{j = 0}^{\rm n} R'_{j,p}) \chf\ome^{(n-1)}_{pj} -\Ps{1}{\f\chi}^n_i h^2  \sum_{j = 0}^{\rm n} R_{j,p}\chf\alp^{(n-1)}_j \,.	\label{eq:coll_disc_rotaz}
\eEq
 
After the decoupling, we perform a mass scaling for the field equations
\bEq 
 \sum_{j = 0}^{\rm n} R_{j,p} \chf a_j^n  =\fr {\f \psi^n_i}{\mu}\, \sepr{with} i = 1,\ldots,\rm n-1 	\label{eq:coll_disc_tran_v2} 
\eEq
\bEq 
\sum_{j= 0}^{\rm n} R_{j,p}{\Del\chf \alp^n_j} =-{\lf(\fr{\der\fR r_i^n(\hf\alp^n_i)}{\der \f\alp^n_i}\rg)}^{-1} {\hfR r_i^n} \, \sepr{with} i = 1,\ldots,\rm n-1\,,\label{eq:coll_disc_rot_v2} 
\eEq
and for the Neumann boundary equations as well  
\bEq
\sum_{j = 0}^{\rm n} R'_{j,p} \chf a^n_j= {(\Ps{2}{\f\psi}^n_i) }^{-1}\fr{\f F^n_i}{h^2},	\label{eq:coll_disc_tran_Neubc4}\\
\eEq
\bEq
 \sum_{j = 0}^{\rm n} R'_{j,p} \chf\alp^n_j=
\ {(\Ps{2}{\f\chi}^n_i ) }^{-1}\fr{\f C^n_i}{h^2}\,,	\label{eq:coll_disc_rot_Neubc4}
\eEq
where we have set
\bEq
\f F^n_i=\baf\psi^n_i - h (\Ps{1}{\f\psi}^n_i \sum_{j = 0}^{\rm n} R_{j,p} \chf\ome^{(n-1)}_{pj} + \Ps{2}{\f\psi}^n_i \sum_{j = 0}^{\rm n} R'_{j,p}\chf v^{(n-1)}_{pj})-\Ps{1}{\f\psi}^n_i h^2   \sum_{j = 0}^{\rm n} R_{j,p} \chf\alp^{(n-1)}_j,\\
\eEq
\bEq
\f C^n_i=\baf\chi^n_i - h (\Ps{1}{\f\chi}^n_i  \sum_{j = 0}^{\rm n} R_{j,p} + \Ps{2}{\f\chi}^n_i  \sum_{j = 0}^{\rm n} R'_{j,p}) \chf\ome^{(n-1)}_{pj} -\Ps{1}{\f\chi}^n_i h^2  \sum_{j = 0}^{\rm n} R_{j,p}\chf\alp^{(n-1)}_j \,.
\eEq

The predictor--multicorrector approach can be now employed for the dynamics of geometrically exact beams. Moreover, the decoupling between translational and angular accelerations allows to set up two different linear systems that can be solved separately. Namely, we have 
\bEq
\begin{bmatrix}
\f M_{a}&\f 0\\
\f 0&\f M_{\alp}\\
\end{bmatrix} \begin{bmatrix}
\chf a\\
\chf \alp\\
\end{bmatrix}= 
\begin{bmatrix}
\f b_{a}\\
\f b_{\alp}\\
\end{bmatrix}\,. 
\eEq

Consider for example a case of Neumann boundary conditions at both ends of the beam, the system of the translational balance equations, $\f M_{a}\chf a=\f b_{a}$, reads
\bEq
\begin{bmatrix}
\f R'(u^c_0)_{0,p} &\f R'(u^c_0)_{1,p} &\cdots&\f R'(u^c_0)_{\textnormal{n},p} &\\
\f R(u^c_1)_{0,p} &\f R(u^c_1)_{1,p} &\cdots&\f R(u^c_1)_{\textnormal{n},p} &\\
\vdots & \vdots &\ddots& \vdots &\\
\f R(u^c_{\textnormal{n}-1})_{0,p} &\f  R(u^c_{\textnormal{n}-1})_{1,p} &\cdots& \f R(u^c_{\textnormal{n}-1})_{\textnormal{n},p} &\\
\f R'(u^c_\textnormal{n})_{0,p} &\f  R'(u^c_\textnormal{n})_{1,p} &\cdots& \f R'(u^c_{\rm n})_{\textnormal{n},p} &\\
\end{bmatrix} \cdot \begin{bmatrix}
\chf a_0^n  \\
\chf a_1^n   \\
\vdots \\
\chf a_\textnormal{n-1}^n  \\
\chf a_\textnormal{n}^n   \\
\end{bmatrix}= \begin{bmatrix}
{(\Ps{2}{\f\psi}^n_0) }^{-1}{\f F^n_0}/{h^2}  \\
{\f \psi^n_1}/{\mu}   \\
\vdots \\
{\f \psi^n_\textnormal{n-1}}/{\mu}\\
{(\Ps{2}{\f\psi}^n_\textnormal{n}) }^{-1}{\f F^n_\textnormal{n}}/{h^2}  \\
\end{bmatrix}
\eEq
where $\f R(u^c_i)_{j,p}=R(u^c_i)_{j,p}\f I $, being $\f I$ the $3\car3$ identity matrix. 

The mass $\f M_a$ and inertia $\f M_\alp$ matrices, apart from the type of boundary conditions (Neumann or Dirichlet), depend on the spatial discretization, in particular on the degree of NURBS/B-Splines, $p$, and on the number of collocation points, $\rm{n}$. In case of homogeneous constraints (such as clamped--clamped, hinged--hinged, or free--free beams) the two matrices are identical, reducing the storing capacity demand.

\subsubsection{Sensitivity of the predictor--multicorrector method}
\begin{figure}
\centering
{
\subfigure[Dirichlet-Dirichlet system.\label{fig:DD}]
{\includegraphics[width=.55\textwidth]{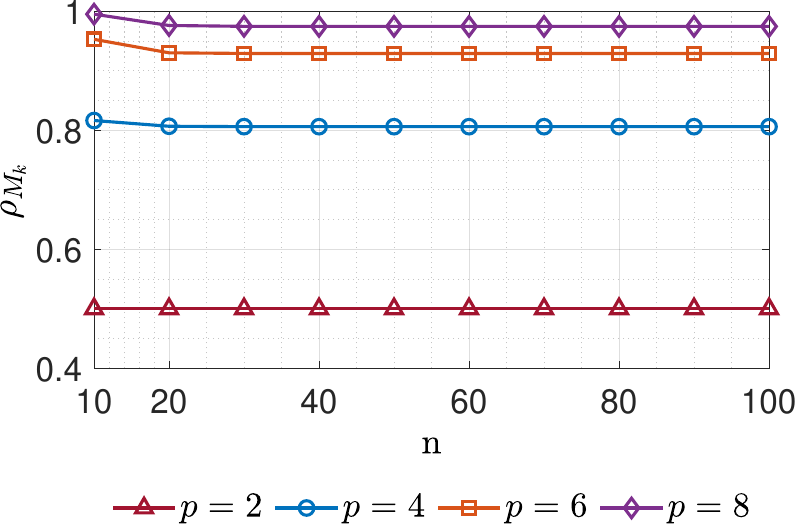}}\hspace{0.5cm}
\subfigure[Dirichlet-Neumann system.\label{fig:DN}]
{\includegraphics[width=.55\textwidth]{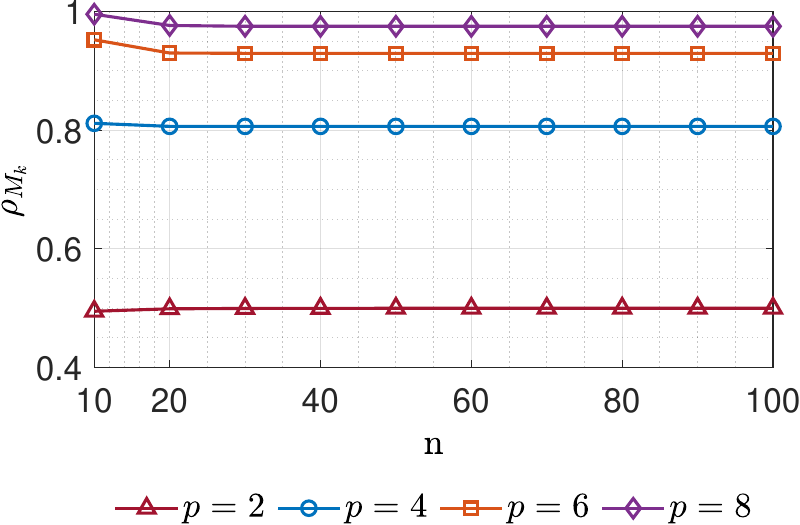}}
\subfigure[Neumann-Neumann system.\label{fig:NN}]
{\includegraphics[width=.55\textwidth]{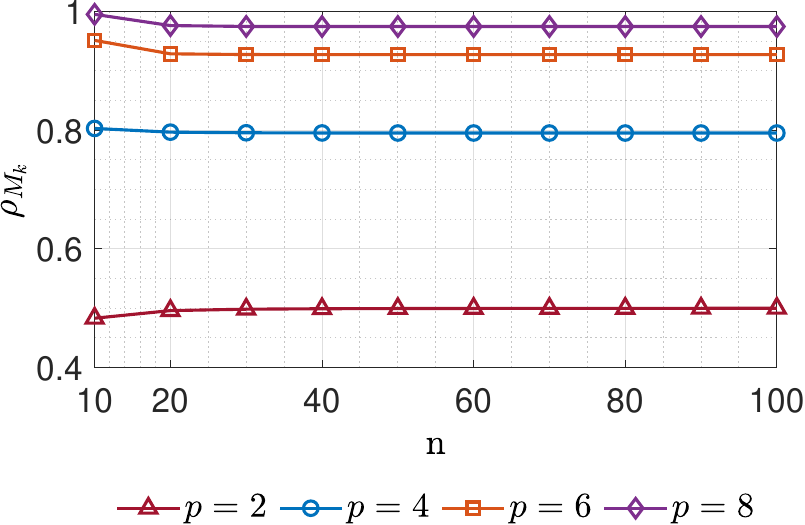}}
\caption{Spectral radius versus number of collocation points for different basis function degree $p$ and different combinations of boundary conditions. Results are the same for both values $k$ can take.}\label{fig:spectral radius}
}
\end{figure}

To check the convergence of the present form of the predictor--multicorrector approach, we study the sensitivity of the spectral radius, $\rho_{\f M_ k} = \rho(\f M_k -\f I) \sepr{with} k = a, \alp$, to variations of $p$, $\rm{n}$, and to the type of boundary conditions. 
We consider three possible combinations of boundary conditions: $i)$ Dirichlet--Dirichlet; $ii)$ Neumann--Neumann and $iii)$ Dirichlet--Neumann. 
Results are presented in Figure~\ref{fig:spectral radius}, where $\rho_{\f M_k}$ is plotted for $p=2, 4, 6, 8$ versus the number of collocation points, $\rm{n}$. Odd degrees are not considered since in collocation they normally present the same convergence rates of smaller even degrees.
We observe that for all the considered combinations, $\rho_{\f M_k}<1$, meaning that convergence is guaranteed. 
Moreover, as already highlighted by \cite{Auricchio2012,Evans_etal2018}, $\rho_{\f M_k}$ tends to increase with $p$ since the band of ${\f M_{k}}$ broadens as the local support of the basis functions becomes wider. On the other hand, different boundary conditions seem to have a negligible impact on $\rho_{\f M_{k}}$.

\subsection{Linear approximation of the rotational balance equation}
As observed in \cite{Marino2019a}, the Newton-Raphson \r{algorithm} normally requires only one iteration. 
This feature suggest that, in an explicit dynamics context where very small \r{time steps are used}, the nonlinearity associated with the angular acceleration term in the rotational balance equation is rather weak. 
On this basis, we explore the appealing possibility to gain a significant advantage in terms of accuracy paying a negligible cost in terms of accuracy. Namely, our assumption is to use directly a linearized form of the rotational balance equation to completely bypass the iterative solution scheme. This is done by substituting one of the two $\f \alp_i^n$ appearing in the left hand side of Eq.~\eqref{eq:ge_mkin_dyn_tdisc3} with $\f \alp_i^{(n-1)}$. Therefore, Eq.~\eqref{eq:ge_mkin_dyn_tdisc3} becomes linear in $\f\alp_i^n$ and reads as follows
\bEq
 \lf[\f j_i^n +\wti{\lf(\fr{h}{2}\f \ome^{(n-1)}_{p,i}+\fr{h^2}{4}\f \alp_i^{(n-1)}\rg)}\f j_i^n\rg]\f \alp_i^n = \f \chi^n_i - [\f \ome^{(n-1)}_{p,i}+\fr{h}{2}\f \alp_i^{(n-1)}]\times \f (j_i^n \f \ome^{(n-1)}_{p,i})\,, \label{eq:ge_mkin_dyn_tdisc4}
\eEq
Eq.~\eqref{eq:ge_mkin_dyn_tdisc4} is then discretized in space and rearranged following the same procedure presented in Section~\ref{PMA}.

To recap, the proposed solution procedure is based on three modifications of the existing formulation: 
i)  the decoupling of the Neumann boundary conditions; 
ii) the rearrangement of the system of equations to obtain a banded mass matrix, $\f M$, containing only basis functions and their derivatives evaluated at the collocation points (see Eq.~\eqref{eq:ge_mkin_dyn_tdisc4}); 
iii) the use of a \r{linearized} form of the rotational balance equation to avoid the iterative scheme. 
We remark that, with these modifications, it is possible to subdivide the $6\rm{n}\car6\rm{n}$ system of equations into two $3\rm{n}\car3\rm{n}$ subsystems, on which the predictor--multicorrector approach can be applied separately. 

\section{Numerical results and discussion\label{sec:num_app}}
In this section, we present the results of the proposed \emp{fully explicit} IGA-C solution procedure, referred to as LU L (LUmped Linear), and compare it with the one in \cite{Marino2019b} employing a consistent mass matrix and solving the original nonlinear system. We refer to the latter consistent nonlinear form as CN NL.
Moreover, to assess our assumption on the linear form of the rotational equation, we provide the results of the predictor--multicorrector approach applied to the nonlinear system of equations (Eqs.~\eqref{eq:coll_disc_translat} and~\eqref{eq:coll_disc_rotaz}). 
We refer to this formulation as LU NL (LUmped NonLinear).  With this additional comparison, we demonstrate that LU L does not introduce any \r{significant} loss of accuracy. Note that, as for CN NL, LU NL still requires the Newton-Raphson scheme with the predictor--multicorrector algorithm applied at each iteration. 

\r{Four} test cases are studied. 
Firstly, a cantilever beam under a constant tip vertical load is analyzed. 
Then, we study a swinging flexible pendulum oscillating under self-weight \r{and a three-dimensional flying beam subjected to tip forces and couples}. The last numerical test concerns a \r{spinning beam in a gravitational field undergoing rigid-body motions}.
The study is completed with a comparative analysis of the computational costs. 

\subsection{Cantilever beam}\label{subsec:cantilever_beam}
\begin{figure}[h]
\centering
\begin{overpic}[width=0.8\textwidth]{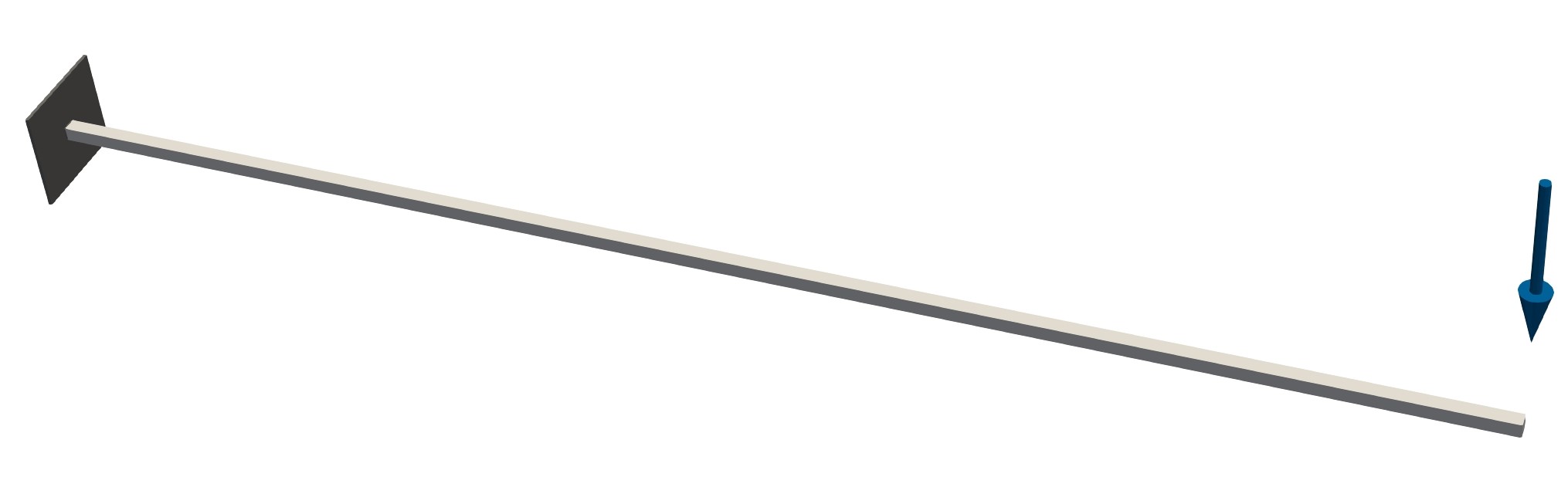}
\put(925,200){$F_3$}
\end{overpic}
\caption{Cantilever beam under impulsive load: geometry and applied forces.\label{fig:cantilever_beam}}
\end{figure}

\begin{figure}
\centering
{\includegraphics[width=1\textwidth]{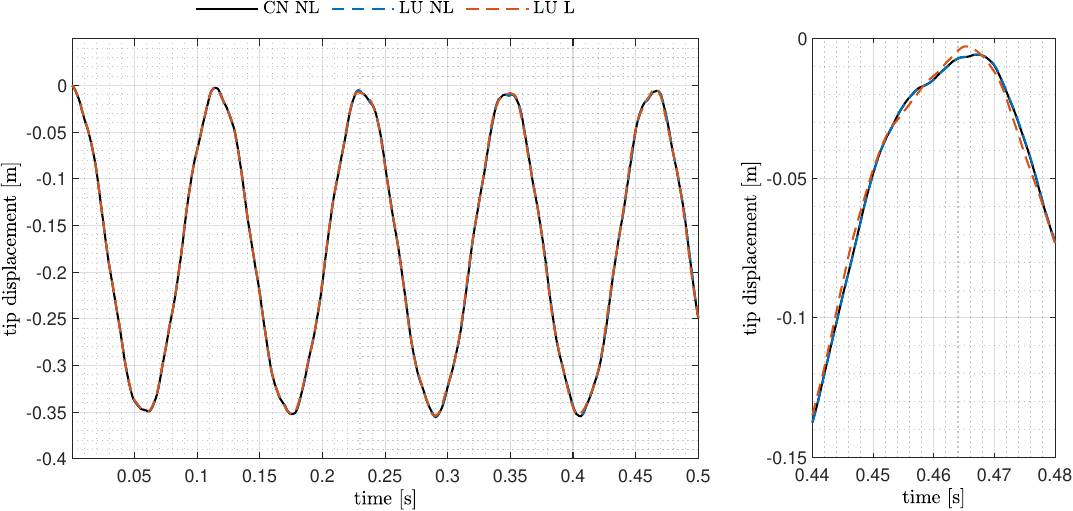}}
\caption{Tip displacement of a cantilever beam subjected to a tip force $F_3 = \SI{-100}{N}$.  $p=4$, $\rm n = 20$,  $h = \SI{1E-6}{s}$: results for the entire simulation time (left) and close-up for $t\in[\SI{0.44}{},\SI{0.48}{}]\,\SI{}{s}$ (right).\label{fig:cantilever comparison}}
\end{figure}

This problem consist of a straight, $\SI{1}{m}$-long cantilever beam with a square cross section of side length $\SI{0.01}{m}$ \cite{Gravouil&Combescure2001}. 
The beam lies along the $x_2$-axis and deforms in the $(x_2,x_3)$ plane. It is is clamped at one end and loaded at its free end with a concentrated constant force, along $x_3$, $F_3=\SI{-100}{N}$ (see Figure~\ref{fig:cantilever_beam}). 
The material properties are: density $\rho=\SI{7800}{kg/m^3}$, Young's modulus $E = \SI{210E9}{N/m^2}$, and Poisson's ratio $\nu = 0.2$.
Figure~\ref{fig:cantilever comparison} shows the time history of the tip displacement. 
The black solid line refers to the consistent nonlinear formulation (CN NL) \cite{Marino2019a}, whereas the blue dashed line to the lumped nonlinear (LU NL) formulation, and the orange dashed line to the lumped linear one (LU L). 
Overall, a very good agreement is observed. 
At the beginning of the simulation the three formulations are almost identical. As time goes (see, e.g., the peak at ca \SI{0.465}{s}), a slight difference is observed in the LU L. 

\begin{figure}
\centering
{
\subfigure[CN NL. \label{fig:CN_NL_clamp}]
{\includegraphics[width=.3\textwidth]{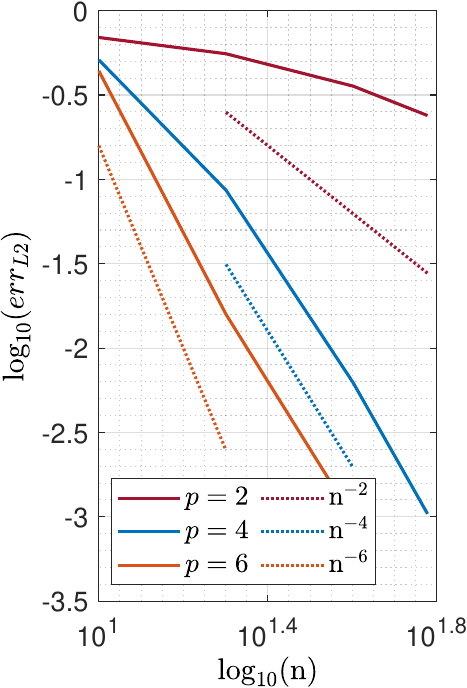}}
\subfigure[LU NL. \label{fig:LU_NL_clamp}]
{\includegraphics[width=.3\textwidth]{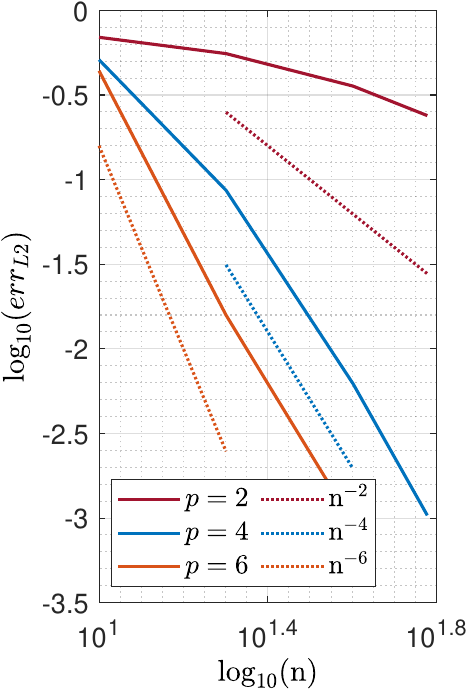}}
\subfigure[LU L.\label{fig:LU_L_clamp}]
{\includegraphics[width=.3\textwidth]{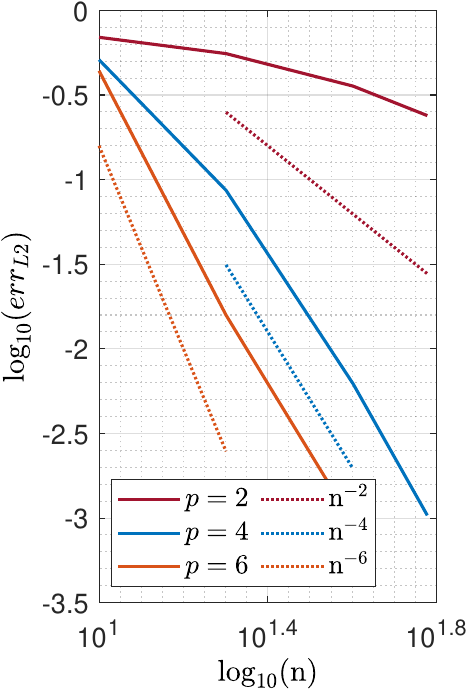}}
\caption{Cantilever beam under tip load: convergence plots for $p=2,4,6$ vs. number of collocation points of the three solution procedures.}\label{fig:convergence plots}
}
\end{figure}

To assess the spatial accuracy of the proposed methods, we perform a convergence study of the $L_2$-norm of the error, $err_{L_2}=||\f{u}^h-\f{u}^r||_{L_2}/||\f{u}^r||_{L_2}$, where $\f u^h$ and $\f u^r$ are the approximated and reference displacements, respectively, evaluated at $t = \SI{0.001}{s}$ and computed over a fixed grid of equally spaced points. 
For each approach, the reference solution is computed with $p=6$, $\rm n=80$ and $h=\SI{1e-7}{s}$. 
In order to minimize the effects of the temporal error, the time step size is reduced when $\rm n$ and $p$ are increased \cite{Marino2019b}. 
The convergence curves are shown in Figure~\ref{fig:convergence plots}. 
The CN NL case is presented in Figure~\ref{fig:convergence plots}a, whereas LU NL and LU L in Figure~\ref{fig:convergence plots}b and  ~\ref{fig:convergence plots}c, respectively. No differences are observed in the rates among the three approaches, demonstrating the capability of the proposed fully explicit method, LU L, to keep the same IGA-C high-order space accuracy of the reference formulation CN NL. 
Concerning $p=2$, it shows a slower convergence as documented also in \cite{Marino2019a,Marino2019b}. 

\subsection{Swinging flexible pendulum}
\begin{figure}[h]
\centering
\begin{overpic}[width=0.8\textwidth]{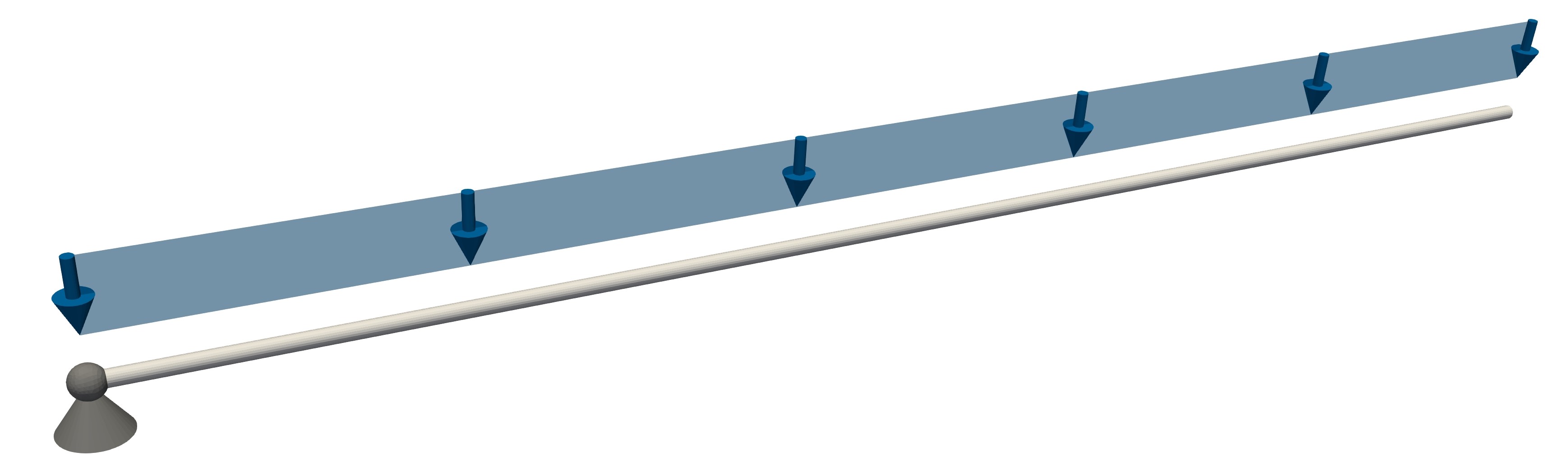}
\put(1000,250){$q_3$}
\end{overpic}
\caption{Swinging flexible pendulum subjected to a distributed vertical load.\label{fig:swing_pendulum}}
\end{figure}

This test concerns a pendulum swinging under the action of its self-weight (see Figure~\ref{fig:swing_pendulum}). The beam presents the same initial geometry of the cantilever case (see Section \ref{subsec:cantilever_beam}), but it is hinged at one end and has a circular cross-section of diameter $\SI{0.01}{m}$ \cite{Weeger_etal2017,Lang_etal2011,Raknes_etal2013,Maurin_etal2015}. The material properties, the same as in \cite{Marino2019b}, are $E = \SI{5E6}{N/m^2}$, $\nu = 0.5$, and $\rho = \SI{1100}{kg/m^3}$. 
The spatial discretization relies on basis functions of degree $p=4$ and $\rm n =30$ collocation points. The simulation lasts $\SI{1}{s}$ with a time step span $h=\SI{1e-5}{s}$. 

\begin{figure}
\centering
\includegraphics[width=1\textwidth]{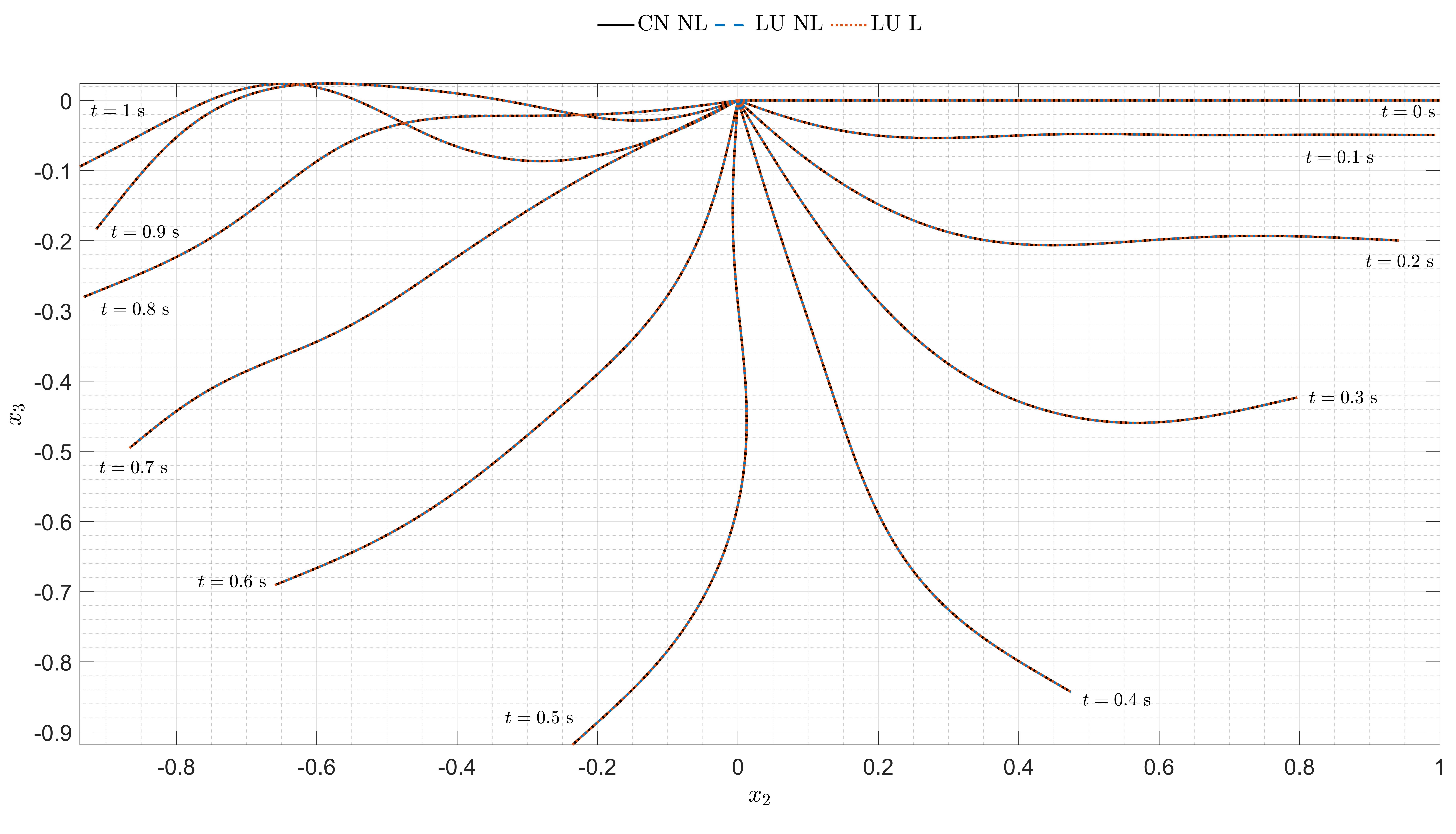}
\caption{Snapshots of a swinging flexible pendulum from time $0$ to $\SI{1}{s}$ with increments of $\SI{0.1}{s}$. $p=4$, $\rm n = 30$, $h = \SI{1E-5}{s}$.}\label{fig:Pendulum_NLNL_configurations}
\end{figure}

\begin{figure}
\centering
\includegraphics[width=0.95\textwidth]{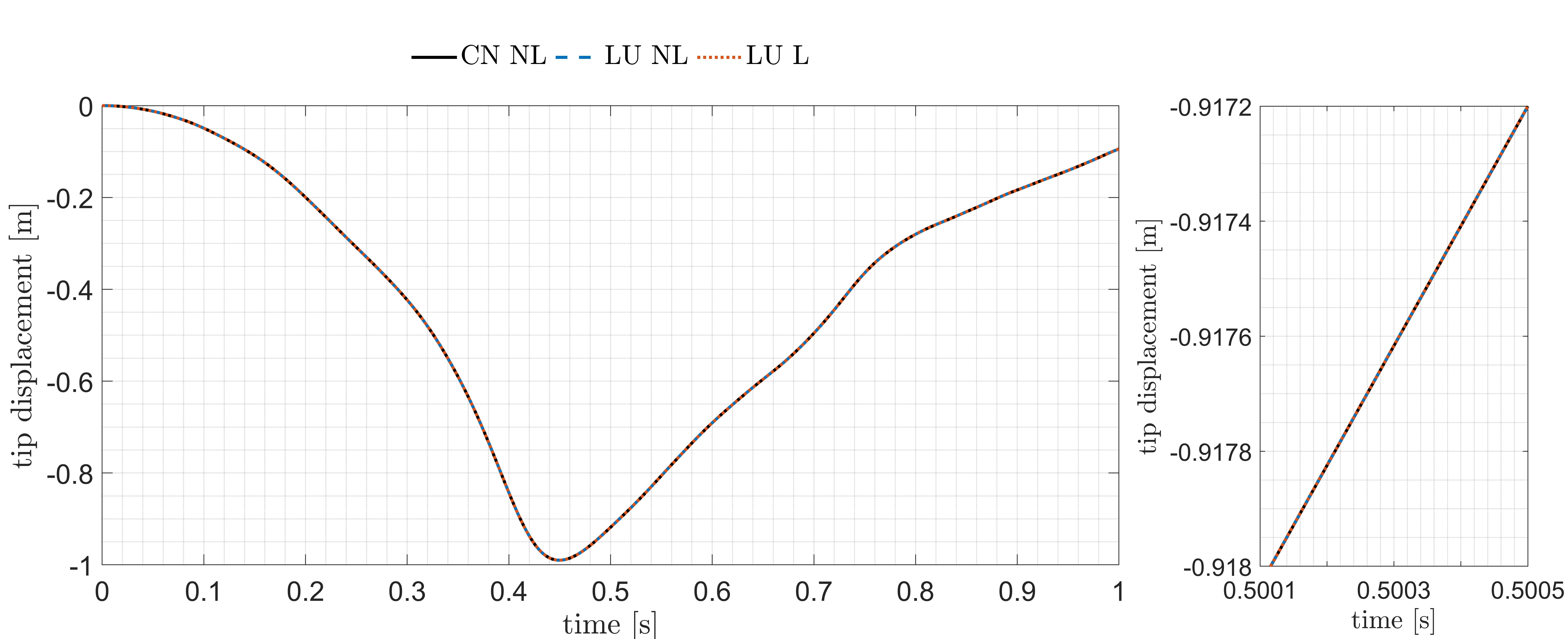}
\caption{Swinging flexible pendulum  results for  $p=4$, $\rm n = 30$, $h = \SI{1E-5}{s}$: tip vertical displacement time history (left) and close up for $t\in[\SI{0.5001}{},\SI{0.5005}{}]\,\SI{}{s}$ (right).}\label{fig:Pendulum_NLNL_Tsim1}
\end{figure}

Figure~\ref{fig:Pendulum_NLNL_configurations} shows some snapshots of the swinging beam. The time history of the tip displacement is shown in Figure~\ref{fig:Pendulum_NLNL_Tsim1}. 
An excellent agreement is observed for all cases.

\begin{figure}
\centering
{
\subfigure[CN NL. \label{fig:CN_NL_pend}]
{\includegraphics[width=.3\textwidth]{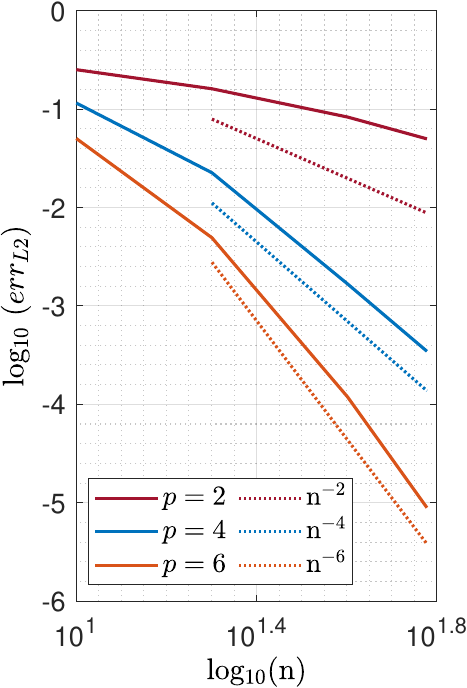}}
\subfigure[LU NL. \label{fig:LU_NL_pend}]
{\includegraphics[width=.3\textwidth]{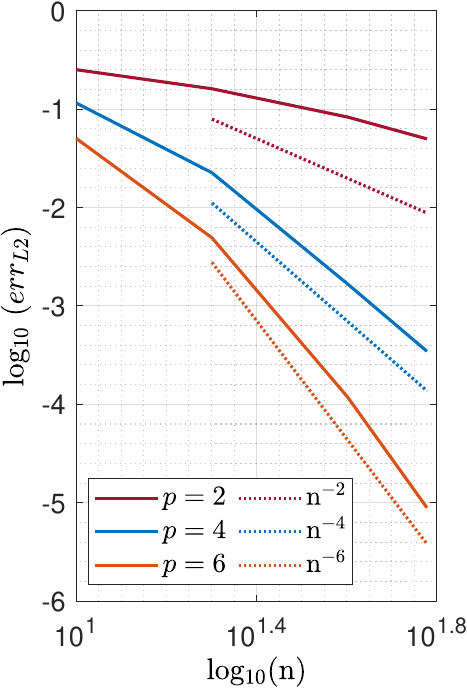}}
\subfigure[LU L.\label{fig:LU_L_pend}]
{\includegraphics[width=.3\textwidth]{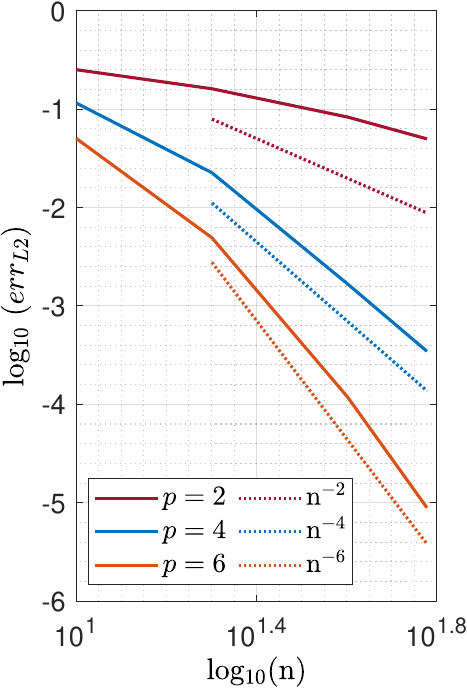}}
\caption{Swinging flexible pendulum: convergence plots for $p=2,4,6$ vs. number of collocation points of the three solution procedures.}\label{fig:convergence plots pend}
}
\end{figure}

To verify the high order spatial accuracy of the proposed method, convergence plots are shown in Figure~\ref{fig:convergence plots pend}.
The $L_2$-norm of the error, $err_{L_2}$, is computed on the displacements evaluated  at $t = \SI{0.1}{s}$. 
The reference solution is obtained with $p=6$, $\rm{n}=80$ and $h=\SI{2.5e-6}{s}$. The rates of convergence are studied for $p=2,4,6$ with $ \rm{n}=10, 20, 40, 60$ and time step sizes equal to $\SI{5e-5}{}$, $\SI{2.5e-5}{}$, $\SI{1.25e-5}{}$ and $\SI{5e-6}{}$, respectively. 
Compared to the reference case, CN NL, excellent rates are achieved also with LU NL and LU L, proving again the 

\subsection{Three-dimensional free flying beam}
With this numerical example, we test the capabilities of our proposed fully explicit formulation to address the well-known problem of the free flying beam undergoing very large and complex three-dimensional motions and rotations \cite{Simo&Vu-Quoc1988,Zupan_etal2012,Hsiao_etal1999,Zhang&Zhong2016,Marino2019a,Marino2019b}. The same material properties of \cite{Simo&Vu-Quoc1988} are used: $\B C_N =\text{diag}(10000,10000,10000)\SI{}{N}$, $\B C_M =\text{diag}(500,500,500)\SI{}{Nm^2}$, $\f J =\text{diag}(10,10,10)\SI{}{kgm^2}$,  and $\mu=\SI{1}{kg/m}$.

\begin{figure}
\centering
{
\subfigure[Initial configuration and loads.\label{fig:FreeFlyingBeam_sketch_forces}]
{\includegraphics[width=.35\textwidth]{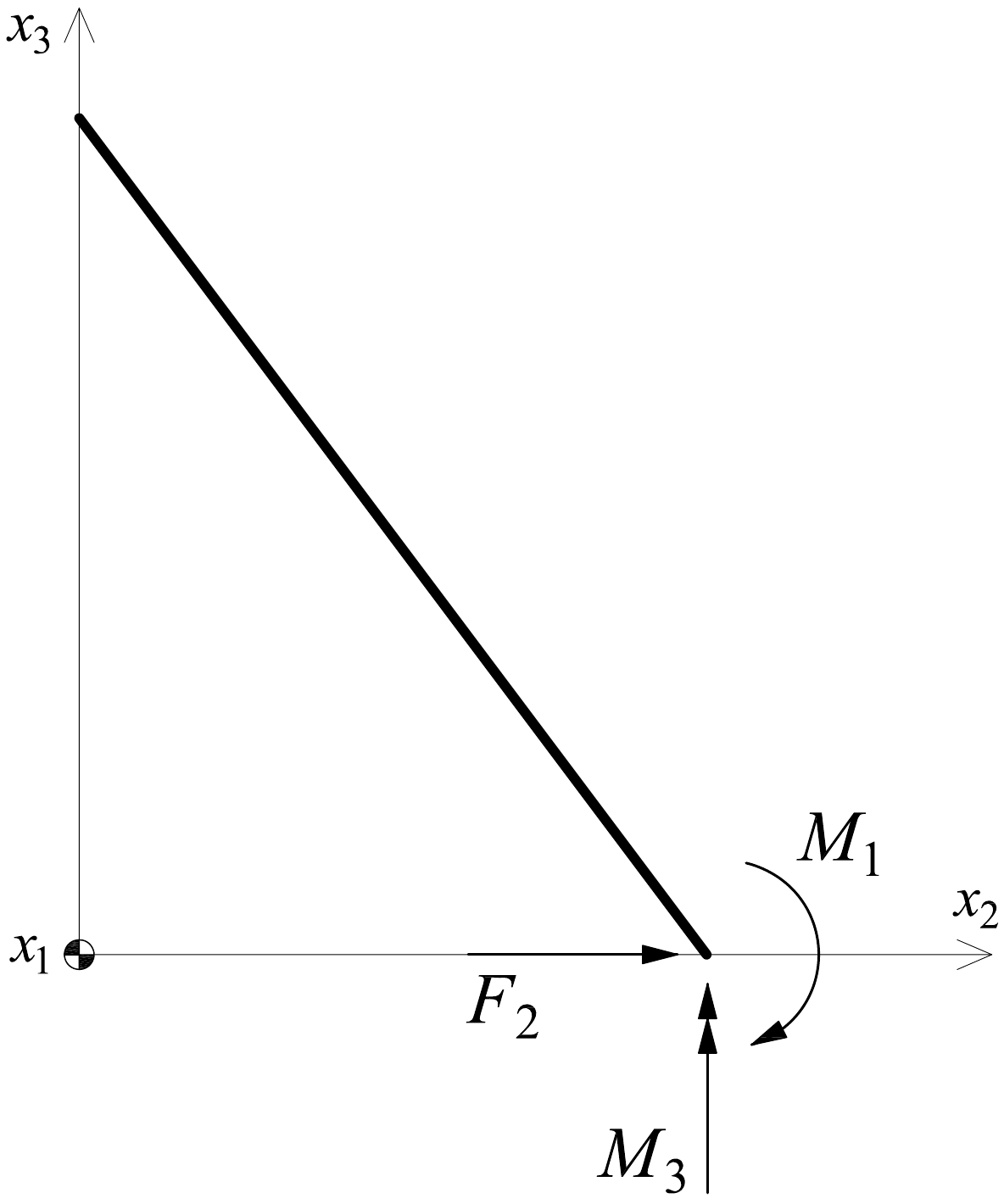}}\!
\subfigure[Load time histories]
{\includegraphics[width=.5\textwidth]{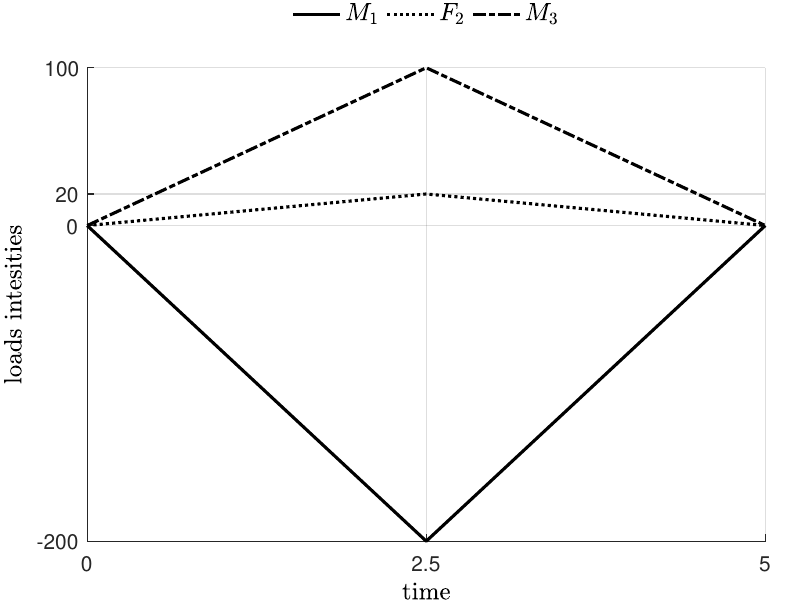}}
}
\caption{Flying flexible beam: initial configuration and loads \cite{Marino2019b}.\label{fig:loads_FlyingFlexBeam3D}}
\end{figure}

Figure~\ref{fig:loads_FlyingFlexBeam3D} shows the initial shape of the beam and the load time histories applied to one of the free ends of the beam. 
The beam axis is discretized with $p=6$ B-Splines and $\rm n = 60$ collocation points. The total simulation time is $T=\SI{5}{s}$ with a time step size of $h=\SI{5e-6}{s}$.

\begin{figure}
\centering
{\includegraphics[width=1\textwidth]{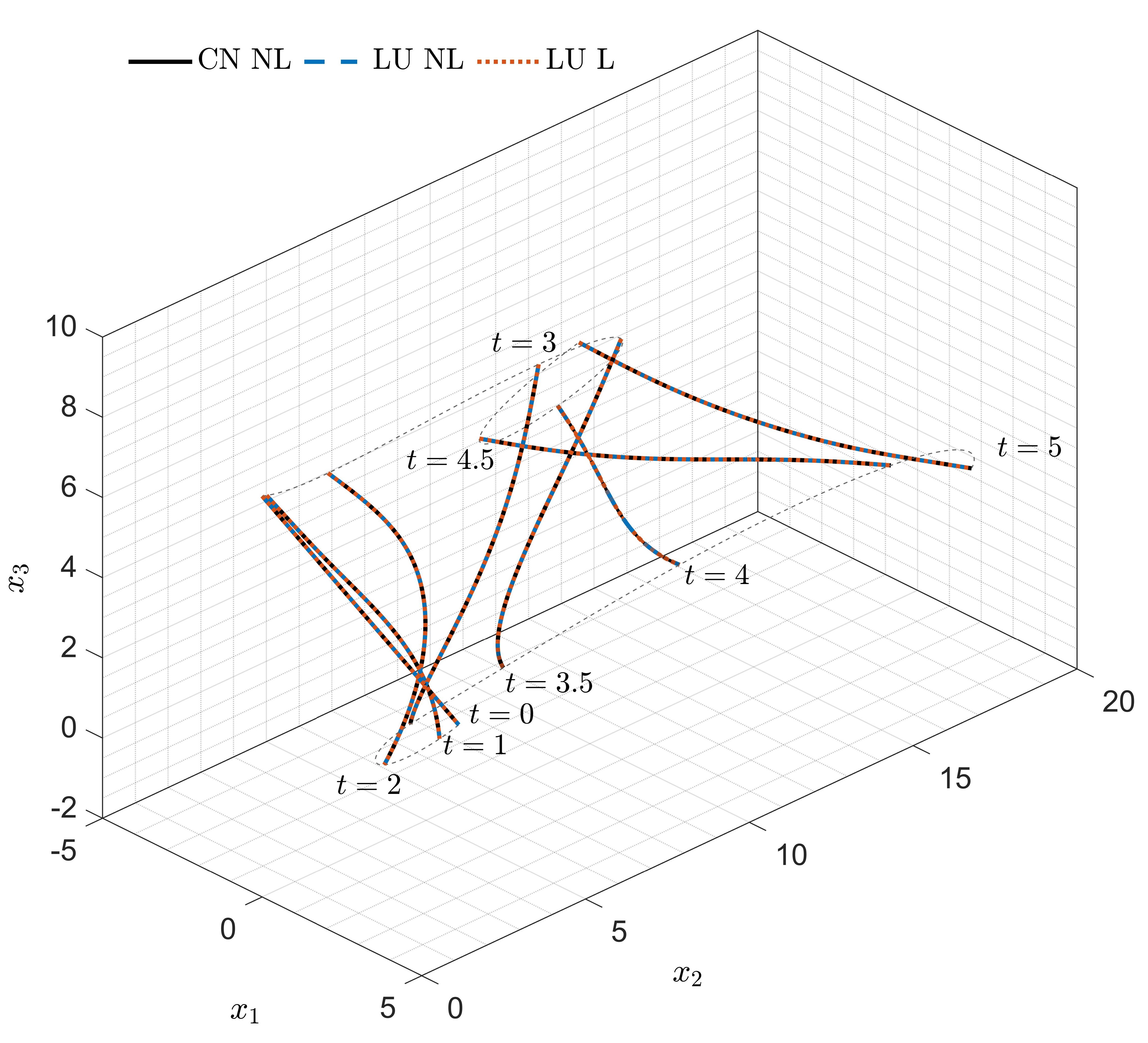}}
\caption{Flying flexible beam: snapshots of the deformed configurations obtained with $p=6$, $\rm n=60$ and $h=\SI{5e-6}{s}$.\label{fig:FreeFlyingBeam_comparison}}
\end{figure}

Snapshots of the deformed configurations obtained with the three different formulations, CN NL, LU NL, and LU L, are plotted in Figure~\ref{fig:FreeFlyingBeam_comparison}. 
No distinguishable differences are observed among the formulations. 

\begin{figure}
\centering
{
\subfigure[CN NL. \label{fig:CN_NL_free}]
{\includegraphics[width=.3\textwidth]{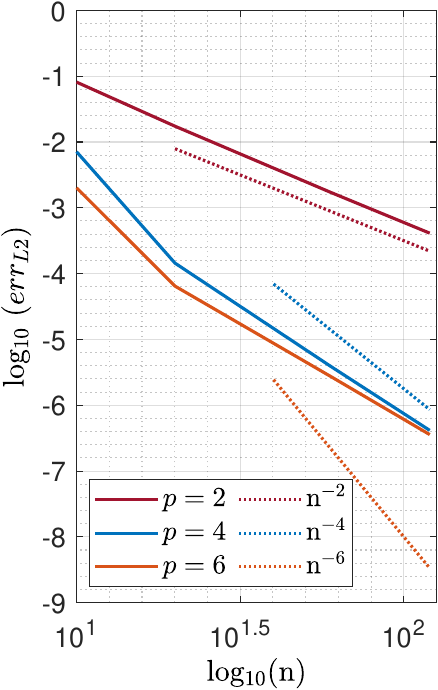}}
\subfigure[LU NL. \label{fig:LU_NL_free}]
{\includegraphics[width=.3\textwidth]{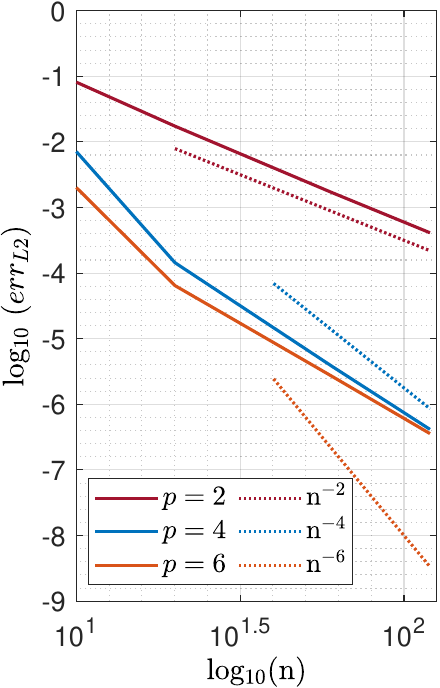}}
\subfigure[LU L.\label{fig:LU_L_free}]
{\includegraphics[width=.3\textwidth]{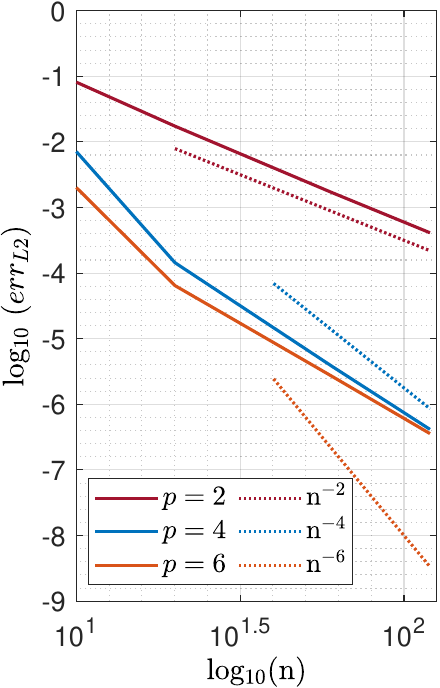}}
\caption{Free flying beam: convergence plots for $p=2,4,6$ vs. number of collocation points of the three solution procedures.}\label{fig:convergence plots free}
}
\end{figure}

The spatial convergence rates of the $L_2$-norm of the error $err_{L_2}=||\f{c}^h-\f{c}^r||_{L_2}/||\f{c}^r||_{L_2}$ computed at $t=\SI{0.5}{s}$ are shown in Figure~\ref{fig:convergence plots free}. $\f c^h$ and $\f c^r$ are the approximate and reference position vectors of the beam centroid, respectively,
The reference solution is obtained with $p=6$ and $\rm{n}=150$, and $h=\SI{5e-6}{s}$.
The dominance of the temporal error over the spatial one is clearly noticeable as the convergence rates do not go beyond the fourth order.
However, we remark that the rates of both LU NL and LU L are identical to the \r{reference formulation} CN NL (Figure~\ref{fig:convergence plots free}a). This clearly indicates that the sub-optimal rates for $p = 6$ are not ascribable neither to the mass lumping nor to the the linearization of the rotational equations, but only to the low (second-) order accuracy of the time integrator.  

\r{
\subsection{Spinning beam in a gravitational field}
In this last numerical example we consider a spinning beam in a gravitational field. 
This test is carried out to further test the capabilities of the proposed method to address large and mainly rigid-body rotations, since the elastic deformations are very small.
The same beam considered in Section 4.1, but with a different square cross-section of side length $\SI{0.0175}{m}$, is hinged at one of its end and loaded by its self-weight, $q_3$ (see Figure~\ref{fig:spin_beam}). Additionally, we prescribe an initial angular velocity to each beam point, $\f\ome^0=[0, 0 ,  \ome^0_{3}]\Tra$. 
We select three different values for $\ome^0_{3}$: 
i) $\ome^0_{3}=0.2\pi$, reproducing a very slow spinning beam dominated by the gravitational load and leading to large three-dimensional motions;
ii) $\ome^0_{3}=2\pi$, where we have a combination of fast rotations and out-of-plane deflections;
iii) $\ome^0_{3}=20\pi$, where the angular velocity is sufficiently high to keep the rigid-body motion almost entirely in the plane $(x_1,x_2)$. 
\begin{figure}[h]
\centering
\begin{overpic}[width=0.6\textwidth]{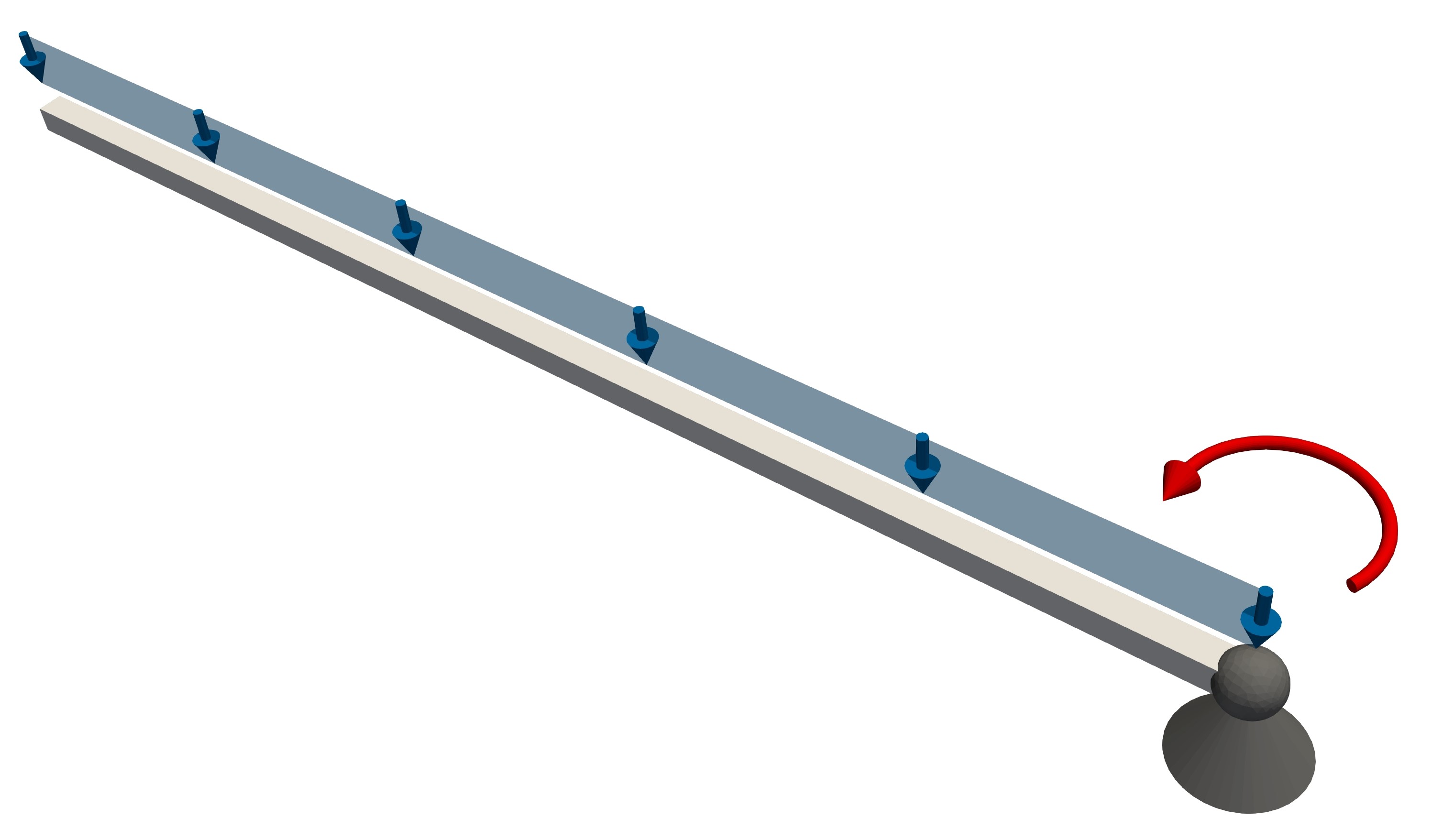}
\put(-30,575){$q_3$}   \put(975,270){$\f\ome^0$}
\end{overpic}
\caption{Spinning beam under gravitational load.\label{fig:spin_beam}}
\end{figure}

Snapshots of the beam motion and tip displacement time histories are reported in Figures~\ref{fig:spin_slow},~\ref{fig:spin_medium} and~\ref{fig:spin_fast} for $\ome^0_{3}=0.2\pi,\,2\pi$ and $20\pi$, respectively. Compared to the reference formulation CN NL, excellent results are obtained by both LU NL and LU L lumping schemes.   

\begin{figure}
\centering
\subfigure[Snapshots of the beam motion obtained with $p=4$, $\rm{n}=20$ and $h=\SI{1e-6}{s}$.\label{fig:spin_slow_3D}]
{\includegraphics[width=1\textwidth]{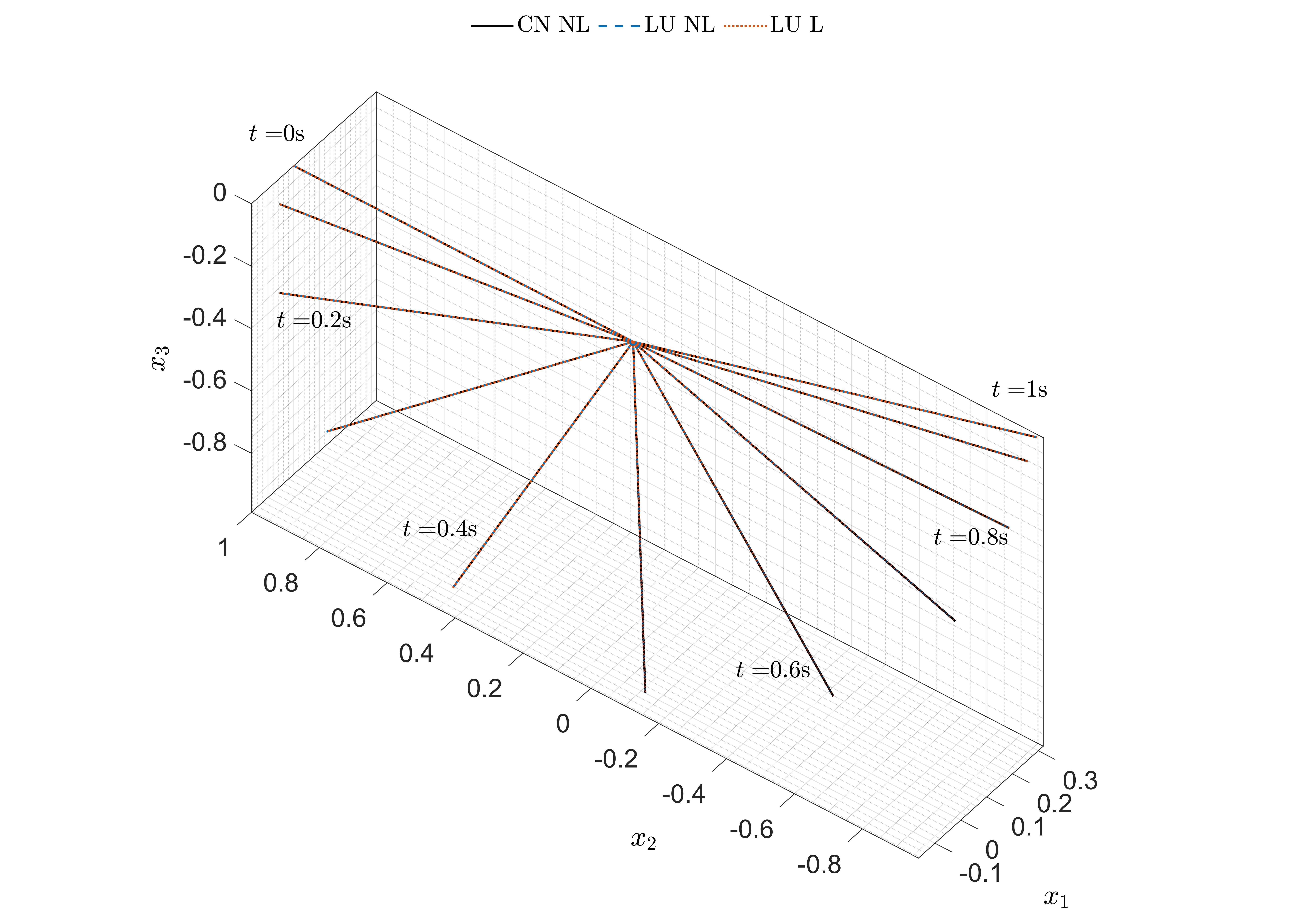}}\\
\bigskip
\subfigure[Tip displacement time history along the $x_1$-axis. \label{fig:spin_slow_u1}]
{\includegraphics[width=0.3\textwidth]{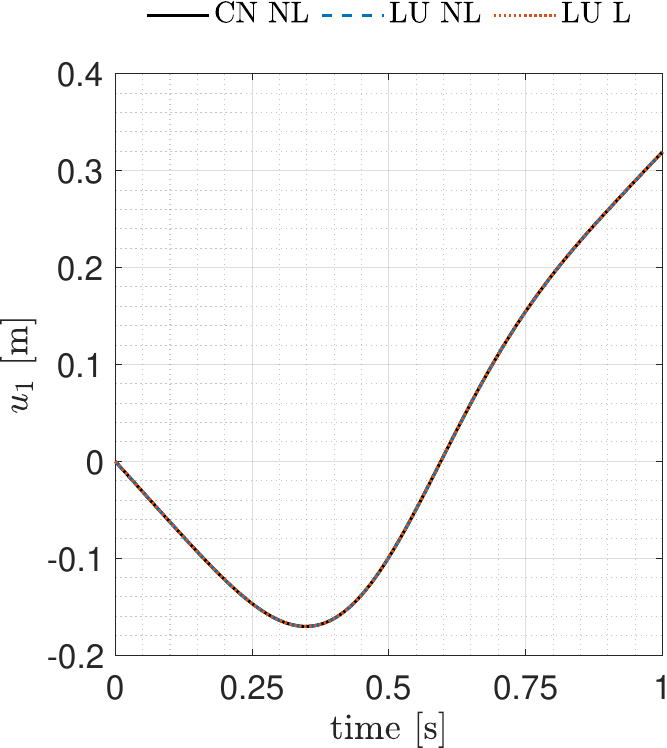}} \hspace{0.3cm}
\subfigure[Tip displacement time history along the $x_2$-axis. \label{fig:spin_slow_u2}] 
{\includegraphics[width=0.3\textwidth]{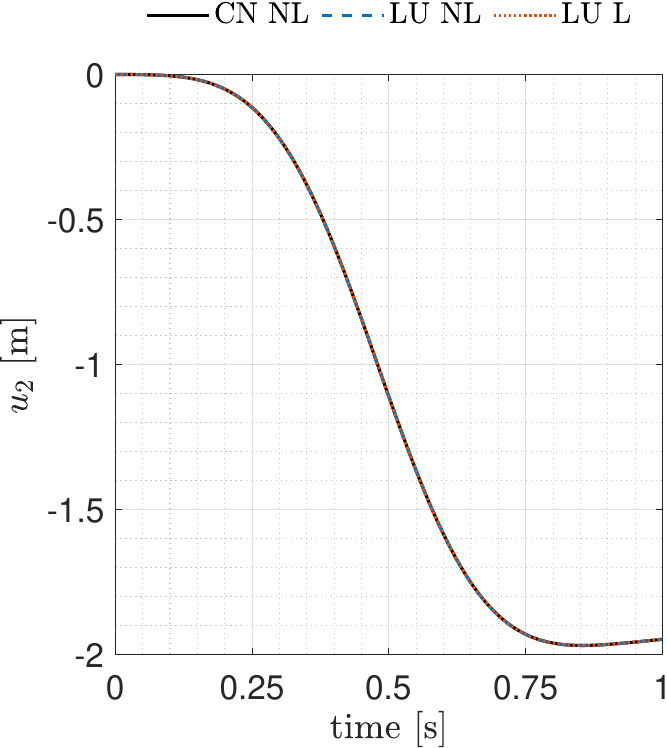}} \hspace{0.3cm}
\subfigure[Tip displacement time history along the $x_3$-axis.\label{fig:spin_slow_u3}]
{\includegraphics[width=0.3\textwidth]{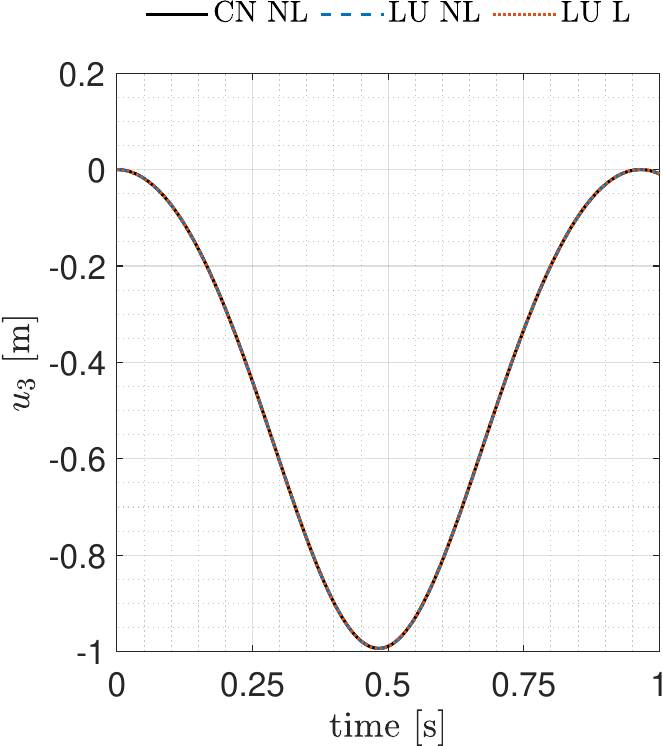}}
\caption{Spinning beam in a gravitational field: results with initial angular velocity, $\ome^0_{3}=0.2\pi$.}\label{fig:spin_slow}
\end{figure}

\begin{figure}
\centering
\subfigure[Snapshots of the beam motion obtained with $p=4$, $\rm{n}=20$ and $h=\SI{1e-6}{s}$.\label{fig:spin_medium_3D}]
{\includegraphics[width=0.9\textwidth]{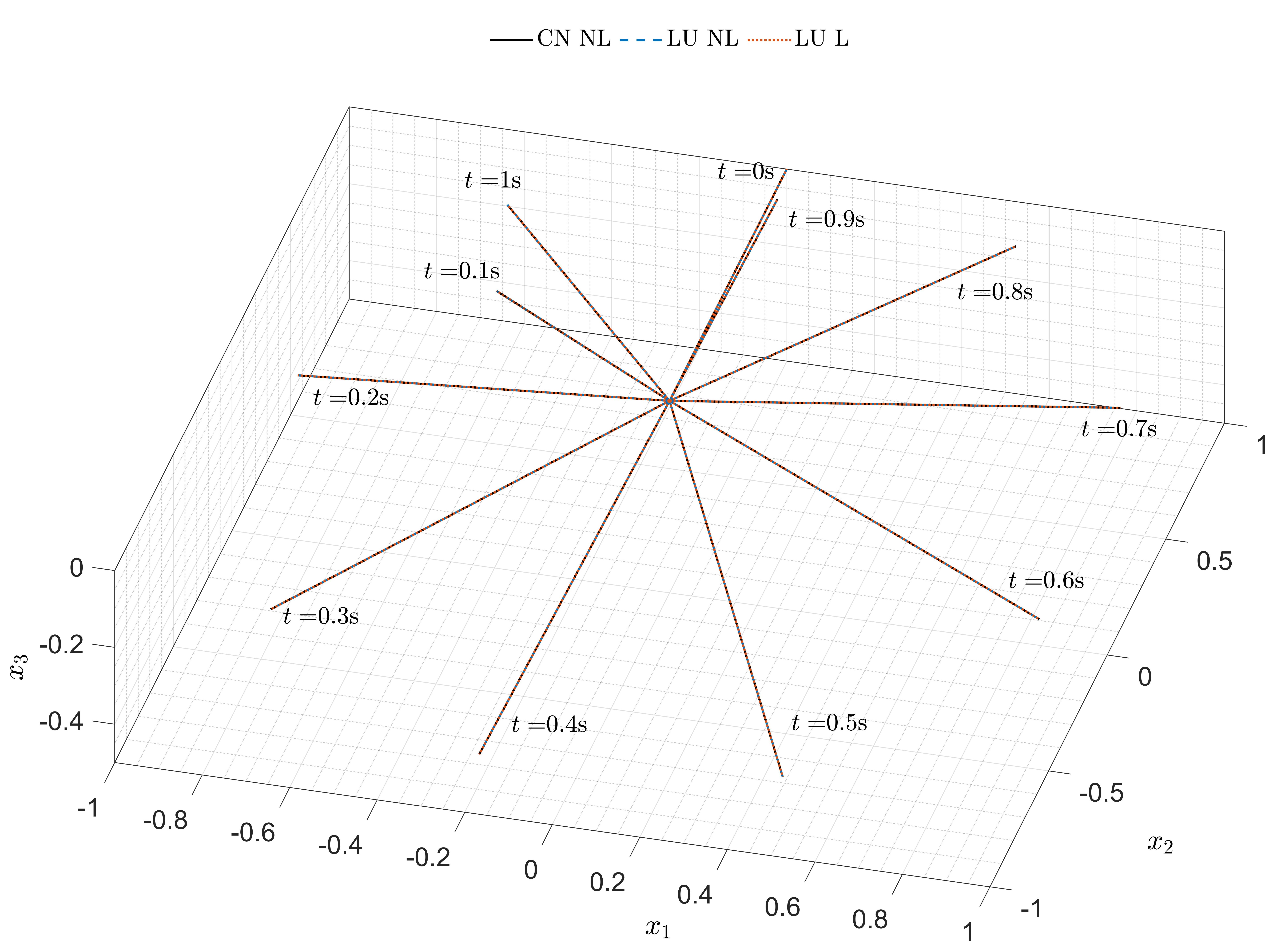}}\\
\bigskip
\subfigure[Tip displacement time history along the $x_1$-axis. \label{fig:spin_medium_u1}]
{\includegraphics[width=0.3\textwidth]{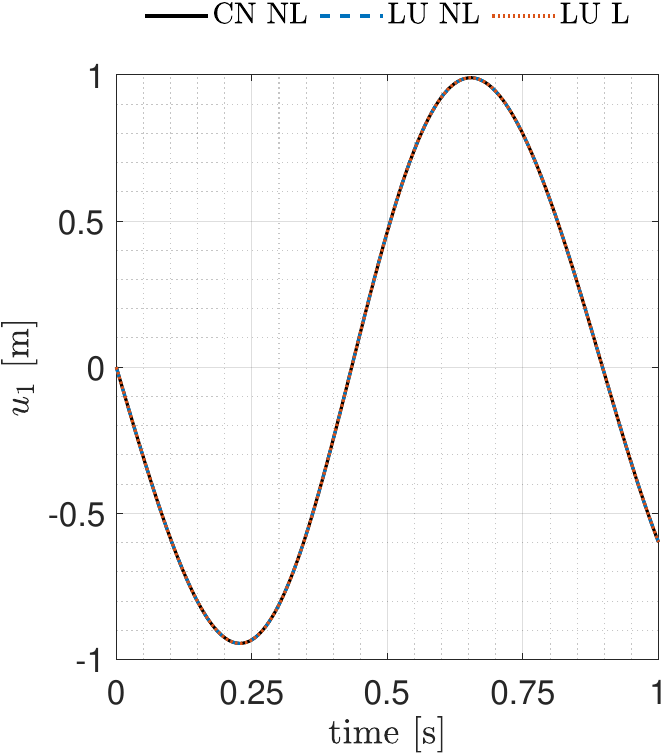}} \hspace{0.3cm}
\subfigure[Tip displacement time history along the $x_2$-axis. \label{fig:spin_medium_u2}] 
{\includegraphics[width=0.3\textwidth]{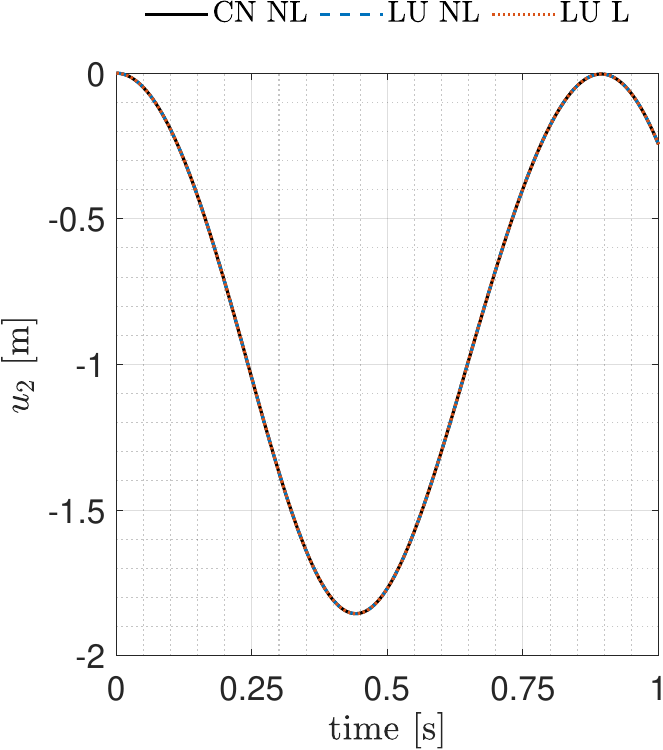}} \hspace{0.3cm}
\subfigure[Tip displacement time history along the $x_3$-axis.\label{fig:spin_medium_u3}]
{\includegraphics[width=0.3\textwidth]{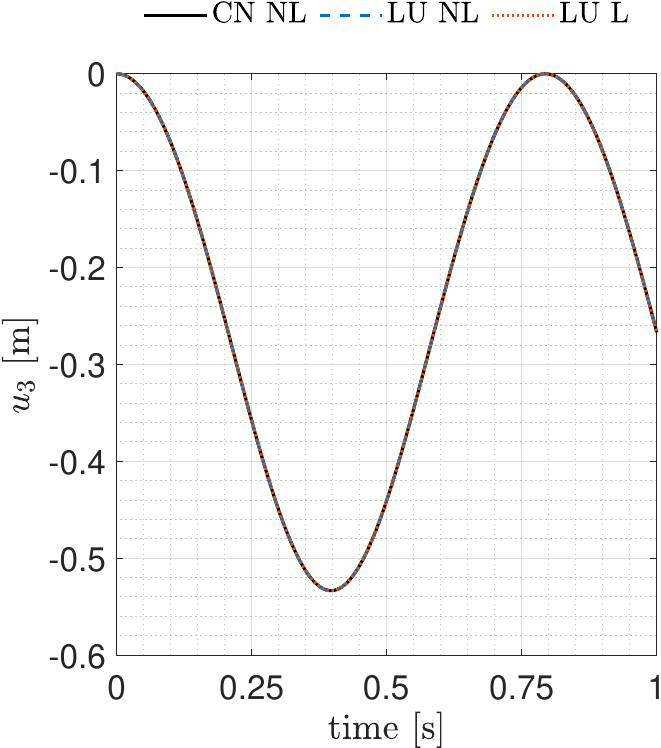}}
\caption{Spinning beam in a gravitational field: results with initial angular velocity $\ome^0_{3}=2\pi$.}\label{fig:spin_medium}
\end{figure}

\begin{figure}
\centering
\subfigure[Snapshots of the beam motion obtained with $p=4$, $\rm{n}=20$ and $h=\SI{1e-6}{s}$.\label{fig:spin_fast_3D}]
{\includegraphics[width=0.9\textwidth]{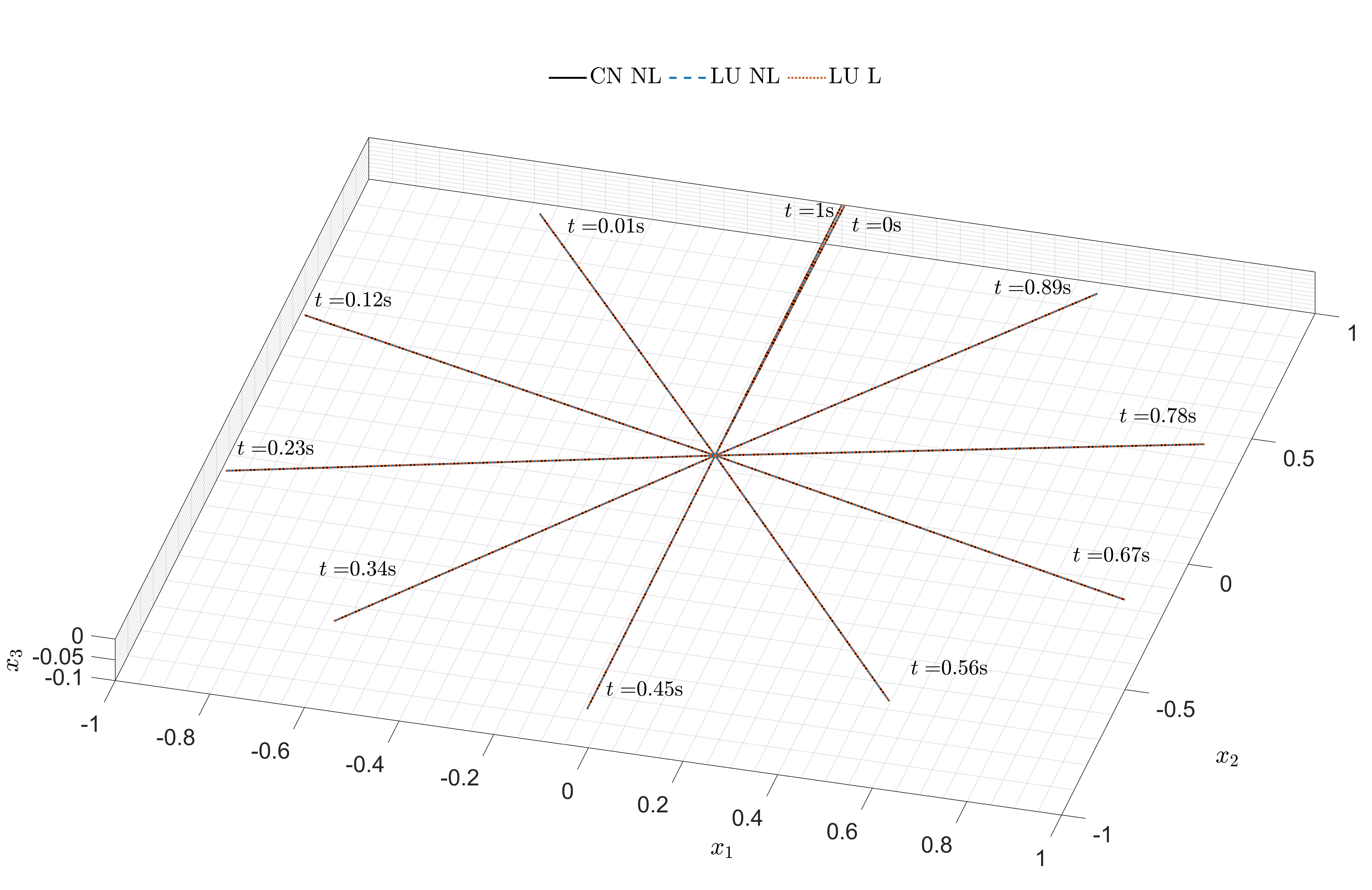}}\\
\bigskip
\subfigure[Tip displacement time history along the $x_1$-axis. \label{fig:spin_fast_u1}]
{\includegraphics[width=0.3\textwidth]{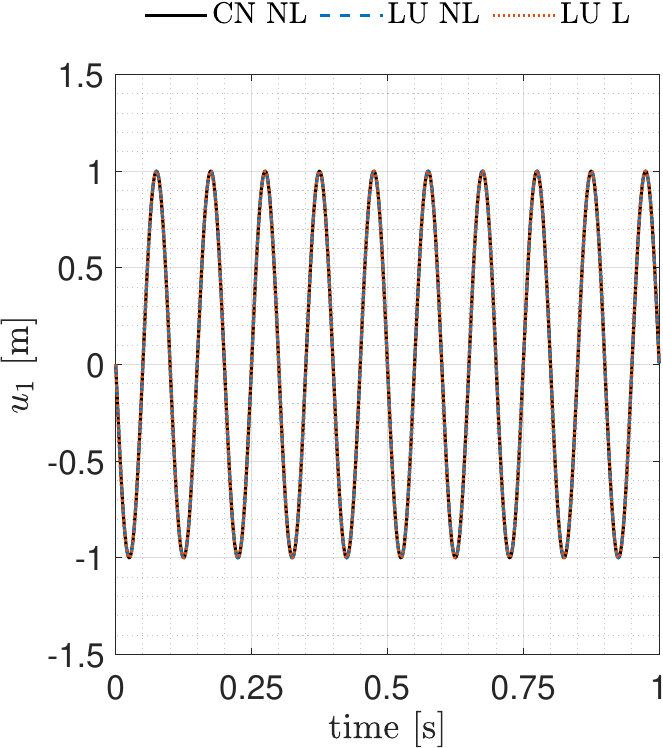}} \hspace{0.3cm}
\subfigure[Tip displacement time history along the $x_2$-axis. \label{fig:spin_fast_u2}] 
{\includegraphics[width=0.3\textwidth]{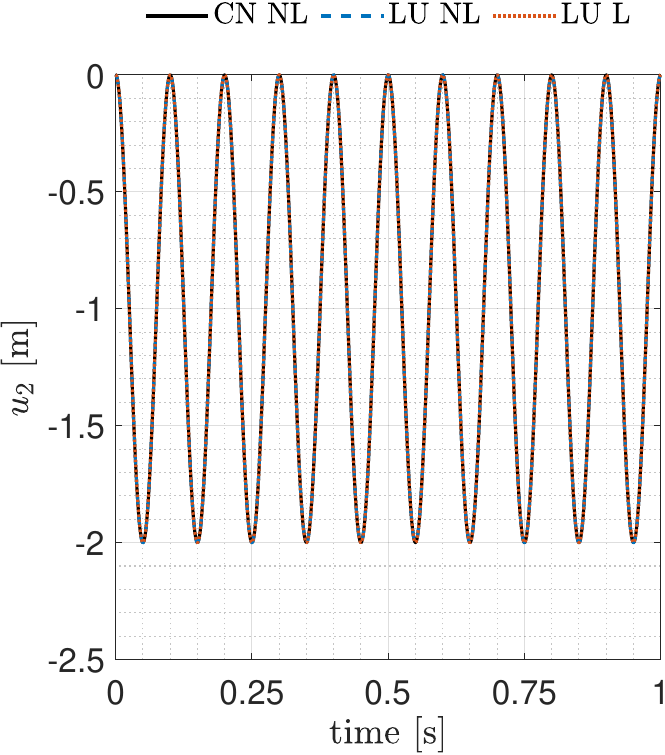}} \hspace{0.3cm}
\subfigure[Tip displacement time history along the $x_3$-axis.\label{fig:spin_fast_u3}]
{\includegraphics[width=0.3\textwidth]{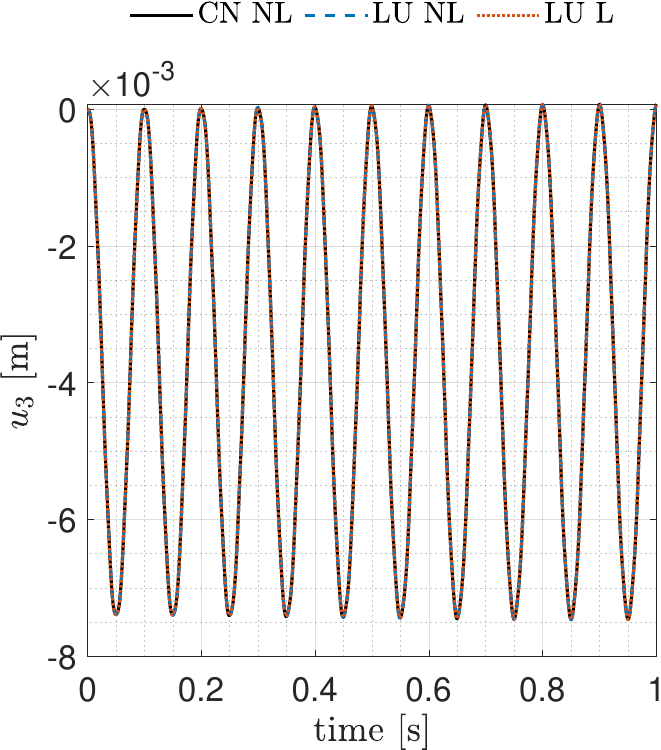}}
\caption{Spinning beam in a gravitational field: results with initial angular velocity, $\ome^0_{3}=20\pi$.}\label{fig:spin_fast}
\end{figure}

}

\subsection{Considerations on the efficiency of the formulations}
In this section, we provide a comparison in terms of computational time among the three formulations: lumped linear, LU L, lumped nonlinear, LU NL, and the reference one, i.e., consistent nonlinear CN NL. 

For each of them, numerical simulations are performed for different values of $p$ and $\rm{n}$, keeping the time size $h$ fixed to the value adopted for the respective overkill solutions. The total time is set to $500h$. The efficiency is estimated calculating the average CPU time required for a single time step, $h_{\rm{CPU}}$. 
Since the scope of this study is to provide quantitative comparisons, rather than to rigorously estimate the computational performances of each formulation, results are plotted in terms of normalized CPU time steps, $h^{{n}}_{\rm{CPU}} = h_{\rm{CPU}}/h^{r}_{\rm{CPU}}$. For a given degree $p$, the reference value, $h^{r}_{\rm{CPU}}$, is taken as the average CPU time per time step obtained with $\rm{n} = 10$ collocation points using the reference CN NL formulation. 

\begin{figure}
\centering
{
\subfigure[$p=2$.\label{fig:CPUtime_p2_clamp}]
{\includegraphics[width=.3\textwidth]{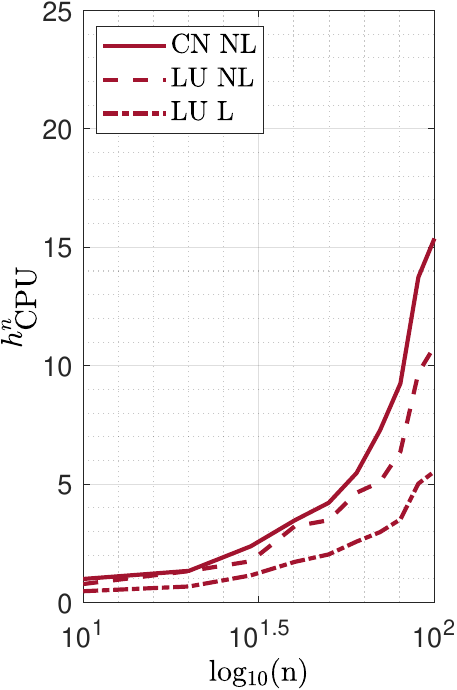}}\hspace{0.3cm}
\subfigure[$p=4$.\label{fig:CPUtime_p4_clamp}]
{\includegraphics[width=.3\textwidth]{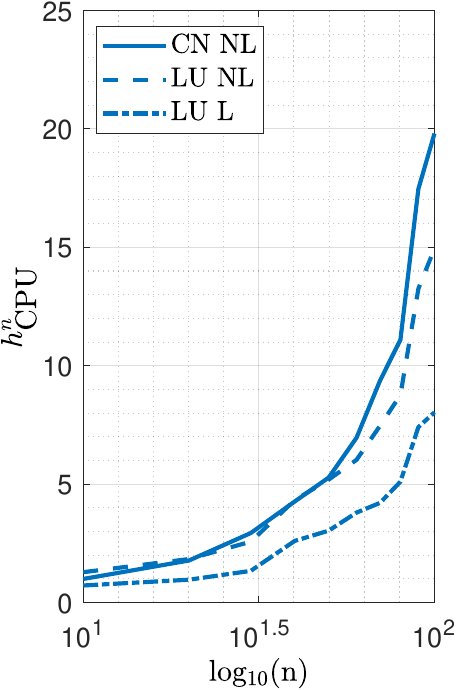}}\hspace{0.3cm}
\subfigure[$p=6$.\label{fig:CPUtime_p6_clamp}]
{\includegraphics[width=.3\textwidth]{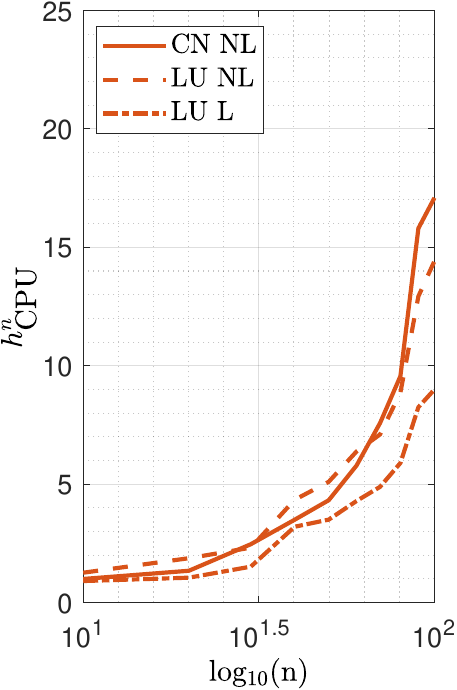}}
\caption{Normalized CPU time per time step vs. number of collocation points for the cantilever beam.}\label{fig:CPU clamp}
}
\end{figure}

\begin{figure}
\centering
{
\subfigure[$p=2$.\label{fig:CPUtime_p2_pend}]
{\includegraphics[width=.3\textwidth]{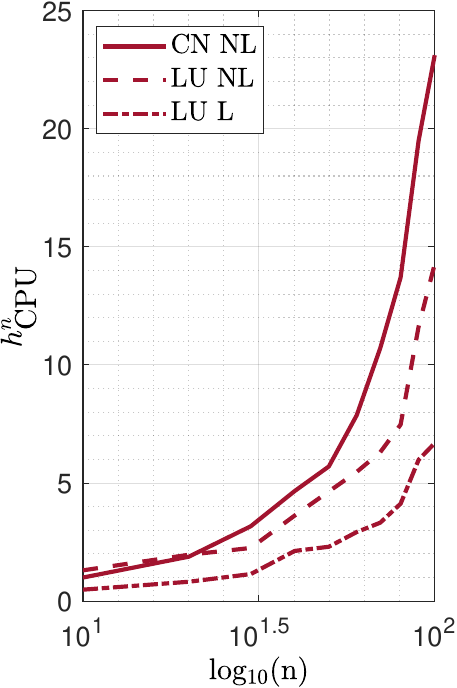}}\hspace{0.3cm}
\subfigure[$p=4$.\label{fig:CPUtime_p4_pend}]
{\includegraphics[width=.3\textwidth]{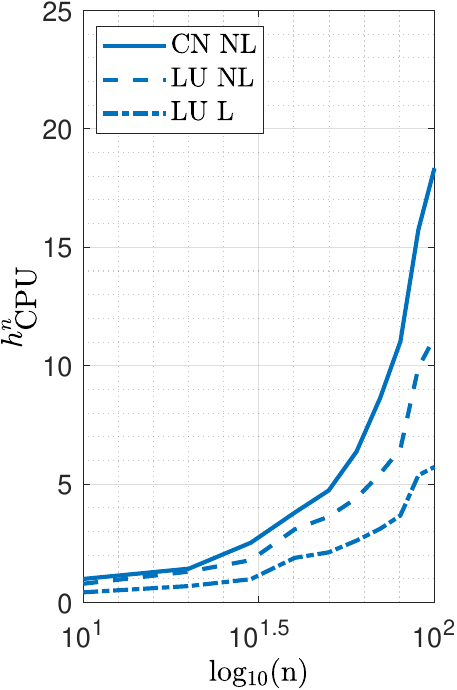}}\hspace{0.3cm}
\subfigure[$p=6$.\label{fig:CPUtime_p6_pend}]
{\includegraphics[width=.3\textwidth]{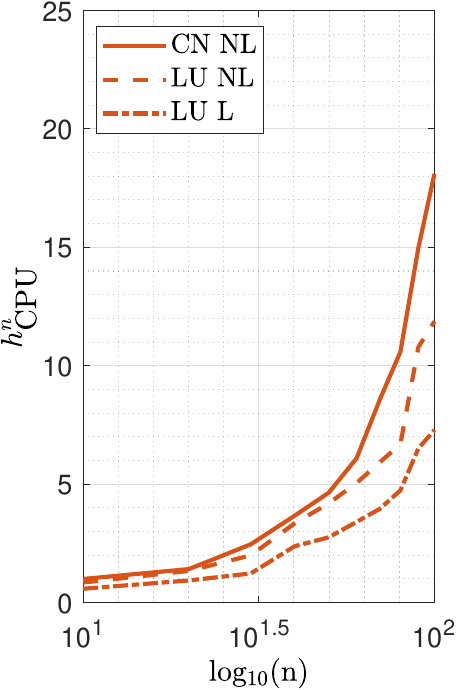}}
\caption{Normalized CPU time per time step vs. number of collocation points for the swinging flexible pendulum.}\label{fig:CPU pend}
}
\end{figure}

\begin{figure}
\centering
{
\subfigure[$p=2$.\label{fig:CPUtime_p2_free}]
{\includegraphics[width=.3\textwidth]{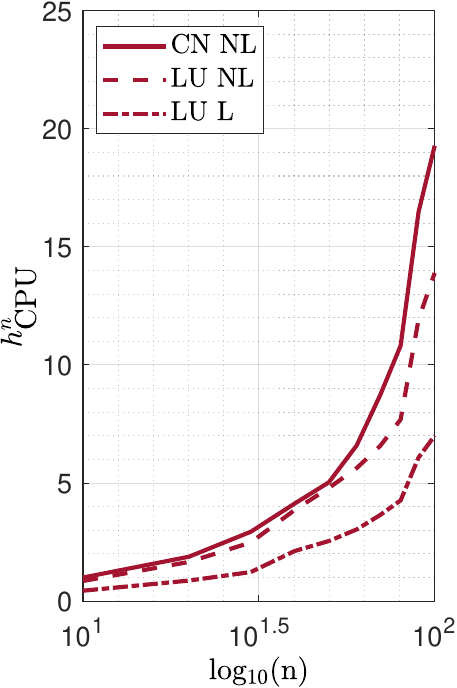}}\hspace{0.3cm}
\subfigure[$p=4$.\label{fig:CPUtime_p4_free}]
{\includegraphics[width=.3\textwidth]{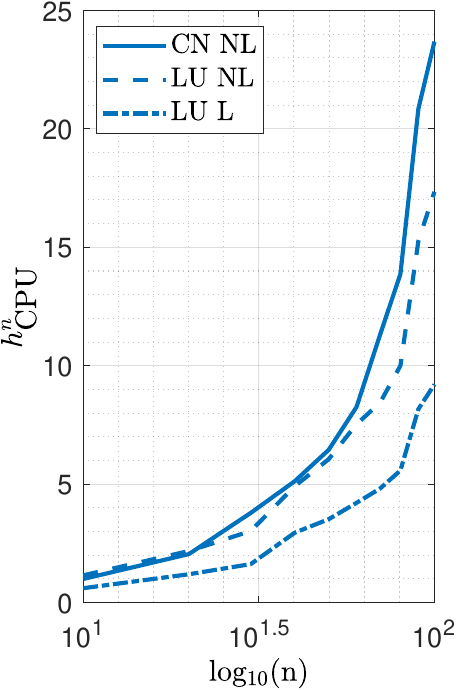}}\hspace{0.3cm}
\subfigure[$p=6$.\label{fig:CPUtime_p6_free}]
{\includegraphics[width=.3\textwidth]{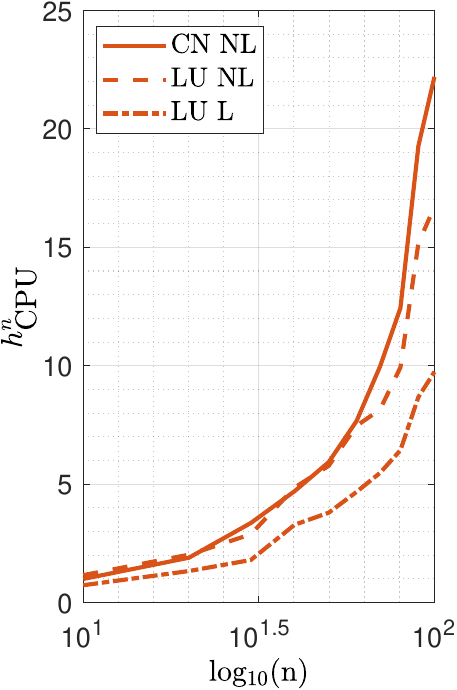}}
\caption{Normalized CPU time per time step vs. number of collocation points for the free flying beam.}\label{fig:CPU free}
}
\end{figure}

\begin{figure}
\centering
{
\subfigure[$p=2$.\label{fig:CPUtime_p2_spin}]
{\includegraphics[width=.3\textwidth]{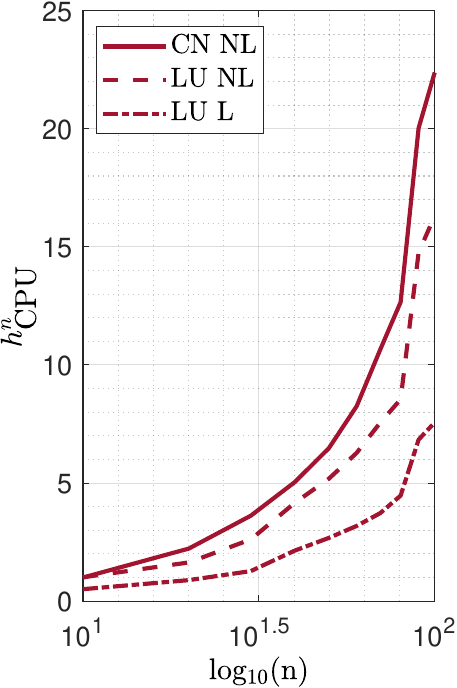}}\hspace{0.3cm}
\subfigure[$p=4$.\label{fig:CPUtime_p4_spin}]
{\includegraphics[width=.3\textwidth]{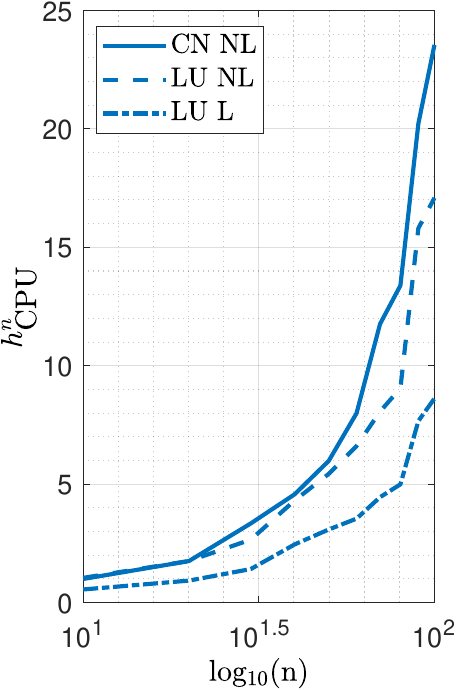}}\hspace{0.3cm}
\subfigure[$p=6$.\label{fig:CPUtime_p6_spin}]
{\includegraphics[width=.3\textwidth]{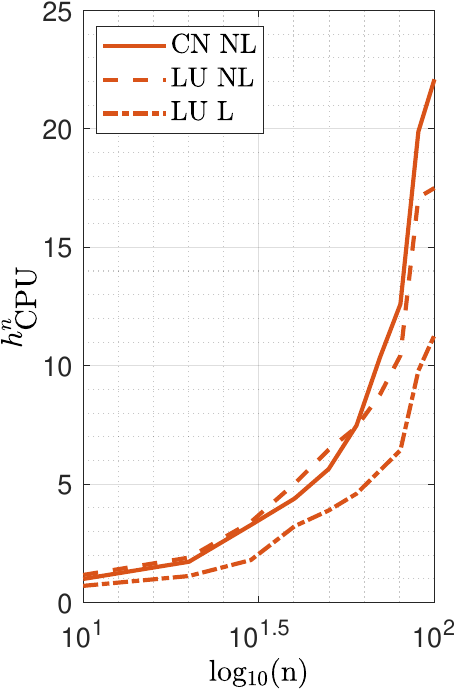}}
\caption{Normalized CPU time per time step vs. number of collocation points for the spinning beam with $\ome^0_{3}=20\pi$.}\label{fig:CPU spin}
}
\end{figure}

Results are presented for the cantilever beam in Figure~\ref{fig:CPU clamp}, for the swinging pendulum in Figure~\ref{fig:CPU pend}, for the flying beam in Figure~\ref{fig:CPU free}, \r{and for the fast spinning beam ($\ome^0_{3}=20\pi$) 
 in Figure~\ref{fig:CPU spin} (similar curves are observed for $\ome^0_{3}=0.2\pi,\,2\pi$, therefore associated plots are not reported)}. 
Solid lines refer to the CN NL case, whereas dashed and dashed dotted lines to LU NL and LU L formulations, respectively. The same colors adopted in the spatial convergence plots (see Figures~\ref{fig:convergence plots},~\ref{fig:convergence plots pend}, and~\ref{fig:convergence plots free}) are here employed for $p=2$ (dark red lines in Figures~\ref{fig:CPU clamp}a--\ref{fig:CPU free}a), $p=4$ (blue lines in Figures~\ref{fig:CPU clamp}b--\ref{fig:CPU free}b),  $p=6$ (orange lines in Figures~\ref{fig:CPU clamp}c--\ref{fig:CPU free}c) curves. 

The LU L formulation exhibits always the lowest $h^{{n}}_{\rm{CPU}}$. In particular, it is noted that the most significant CPU time reduction occurs for large values of $\rm{n}$, meaning that the proposed method has the potential to dramatically increase the efficiency in simulations of complex beams systems with a high number of degrees of freedom, still preserving the high-order accuracy typical of IGA-C. Moreover, except for the clamped case with $p = 4, 6$ (see Figure~\ref{fig:CPU clamp}) \r{and for the spinning beam with $p=6$ (see Figure~\ref{fig:CPU spin})}, the LU NL formulation is faster than the CN NL one, even with a low number of collocation points. 
As expected, as $p$ increases, the efficiency gain tends to reduce due to a larger spectral radius (see Figures~\ref{fig:spectral radius}) which requires more corrector passes.


\section{Conclusions\label{sec:conclusions}}
In this paper, we proposed a fully explicit dynamic IGA-C formulation for geometrically exact beams. Starting from an existing formulation, which is explicit only in the strict sense of the time integration algorithm, we made the method \emp{fully explicit} adapting an existing predictor--multicorrector method, originally proposed for standard linear elastodynamics, to the case of the finite rotational dynamics of geometrically exact beams.

The procedure relies on decoupling the Neumann boundary conditions and on a rearrangement and rescaling of the mass matrix. Moreover, we pursued additional efficiency removing the \r{angular velocity-dependent} nonlinear term in the rotational balance equation, bypassing the need for a time-consuming iterative scheme.
The performance of the method is tested with three numerical applications involving both Dirichlet-Neumann and Neumann-Neumann boundary conditions. 

We demonstrated that the proposed method preserves the same high-order spatial accuracy as the ``exact'' case where a consistent mass matrix and the full nonlinear rotational balance equation are used. 
We also quantified the gain in terms of computational cost and demonstrated that the proposed method significantly decreases the computational time without losing accuracy. This gain increases with the number of collocation points, indicating that the proposed lumping scheme has the potential to manage complex dynamic problems with many degrees of freedom which would be not affordable with methods using the consistent mass matrix.
In some cases, we found that the temporal error dominates the spatial one, regardless of the mass matrix used. Therefore, future works will be devoted to the development of $\SO3$-consistent beam dynamic formulations with higher-order accuracy for both space and time. 
\r{Given the raising interest in the dynamics of multi-body and complex-shaped systems, such as mechanical meta-materials, future developments will also include multi-patch structures.}

\section*{Acknowledgments}
EM was partially supported  by the European Union - Next Generation EU, in the context of The National Recovery and Resilience Plan, Investment 1.5 Ecosystems of Innovation, Project Tuscany Health Ecosystem (THE). (CUP: B83C22003920001).

EM and AR were also partially supported by the National Centre for HPC, Big Data and Quantum Computing funded by the European Union within the Next Generation EU recovery plan. (CUP B83C22002830001). 

EM and GF were partially supported  by the UniFI project IGA4Stent - ``Patient-tailored stents: an innovative computational isogeometric analysis approach for 4D printed shape changing devices''. (CUP B55F21007810001).

JK was partially supported by the European Research Council through the H2020 ERC Consolidator Grant 2019 n. 864482 FDM2.

AR was partially supported by the Italian Ministry of University and Research (MUR) through the PRIN project COSMIC (No. 2022A79M75), funded by the European Union-Next Generation EU.

These supports are gratefully acknowledged.

\newpage
\bibliographystyle{elsarticle-num}
\bibliography{mylibrary_new}

\end{document}